\documentclass[usenatbib]{mnras}
\usepackage{graphicx}

\def \Msun{ {\rm M}_{\sun} }
\def \kms{ {\rm km ~ s^{-1}} }
\def \deg{ ^{\circ} }
\def \kpc{ {\rm kpc} }
\def \bfd{ }
\def \bfr{ }

\title[Classifying S0s with CNNs]{Classifying the formation processes of S0 galaxies using Convolutional Neural Networks}

\author[J. Diaz et al.]
{J.D. Diaz,$^1$\thanks{Email: jonathan.diaz@uwa.edu.au}
    Kenji Bekki,$^1$ Duncan A. Forbes,$^2$ Warrick J. Couch,$^2$
    \newauthor
    Michael J. Drinkwater$^3$, and Simon Deeley$^3$
    \\
    ${}^1$ICRAR, M468, The University of Western Australia 35 Stirling Highway, Crawley Western Australia, 6009, Australia \\
    ${}^2$Centre for Astrophysics \& Supercomputing, Swinburne University, Hawthorn, VIC 3122, Australia \\
    ${}^3$School of Mathematics and Physics, University of Queensland, QLD 4072, Australia \\ 
}

\date{}

\begin{document}

\maketitle
\begin{abstract}

Numerous studies have demonstrated the ability of Convolutional Neural Networks (CNNs) to classify large numbers of galaxies in a manner which mimics the expertise of astronomers.  Such classifications are not always physically motivated, however, such as categorising galaxies by their morphological types.  In this work, we consider the use of CNNs to classify simulated S0 galaxies based on fundamental physical properties.  In particular, we undertake two investigations: (1) the classification of simulated S0 galaxies into three distinct evolutionary paths (isolated, tidal interaction in a group halo, and Spiral-Spiral merger), and (2) the prediction of the mass ratio for the S0s formed via mergers.  {\bfr To train the CNNs, we first run several hundred N-body simulations to model the formation of S0s under idealised conditions; and then we build our training datasets by creating images of stellar density and two dimensional kinematic maps for each simulated S0.}  Our trained networks have remarkable accuracies exceeding 99\% when classifying the S0 {\bfd formation pathway}.  For the case of predicting merger mass ratios, the mean predictions are consistent with the true values to within roughly one standard deviation across the full range of our data.  {\bfd Our work demonstrates the potential of CNNs to classify galaxies by the fundamental physical properties which drive their evolution.}

\end{abstract}

\begin{keywords}
   galaxies: elliptical and lenticular -- galaxies: kinematics and dynamics
\end{keywords}

\section{Introduction} \label{sec:int}

As astronomy enters the era of `big data', scientists will be tasked with extracting knowledge from ever-increasing volumes of data created by facilities such as the LSST and SKA.  To cope with the scale of this task, new techniques are being explored to expand the analytical tools available to scientists.  Among such tools are artificial neural networks, which are a large set of complex algorithms borrowed from adjacent fields such as computer vision.  As an example, numerous recent papers have demonstrated how Convolutional Neural Networks (CNNs) can automate the classification of galaxy morphologies for huge datasets of input images (e.g. \citealt{dieleman2015, ds2018}).

Neural networks are also being developed to supplement or perhaps even supplant the traditional analytical tools for a wide range of astronomical applications.  Such tasks include photometric decomposition of galaxies (e.g. \citealt{tuccillo2018, stark2018}), finding galaxy lenses (\citealt{petrillo2018}), measuring photometric redshifts (\citealt{hoyle2016}), classifying supernovae light curves (e.g. \citealt{charnock2017}), and more.

In essence, these previous efforts demonstrate the potential of various flavours of neural networks to automate and accelerate familiar tasks for astronomers. This is particularly useful not only because ever larger datasets can be passed into the networks for analysis, but also because it enables the scientist to spend more time on scientifically meaningful and challenging tasks.

However, the utility of neural networks (and more broadly, artificial intelligence) in astronomy is not limited to this scope.  In addition to replicating the abilities of human scientists, neural networks may also be used to extend those abilities into new realms by addressing scientific problems which would otherwise lie beyond reach.  {\bfd Recent work in this vein includes the use of CNNs to detect dark matter subhalos by their kinematic imprints in discs (\citealt{aiverse,shah2019}), and using CNNs to constrain the orbits of cluster galaxies from the properties of stripped gas (\citealt{bekki2019}). } In this paper, {\bfd we contribute to this trend} by training CNNs to classify { \bfd S0 galaxies } according to their formation processes.

The morphology and kinematics of present-day galaxies have long been considered to bear imprints of their formation histories (\citealt{buta2007,kormendy2012,forbes2016}).  Extracting this fossil information, however, is a considerable challenge.  For this reason, astronomers have traditionally discussed morphologies in terms of an ad-hoc visual classification known as the Hubble tuning fork (\citealt{hubble36}) and its various revisions (e.g. \citealt{devauc59,sandage61}).

{ \bfd A number of surveys utilising integral field spectroscopy (IFS) have been recently completed or are currently underway, {\bfr including ATLAS3D (\citealt{cap2011}), CALIFA (\citealt{sanchez2012}), SAMI (\citealt{croom2012}), SLUGGS (\citealt{brodie2014}), MASSIVE (\citealt{ma2014}), and MaNGA (\citealt{bundy2015}).}  These surveys have produced a wealth of two dimensional kinematic observations of nearby galaxies and have delivered new insights into galaxy evolution.  The availability of this high resolution kinematic data presents an opportunity to explore new methods for further analysis.  In this work, we consider the possibility that CNNs can leverage the information contained within such kinematic data for the purposes of galaxy classification.  The theme of re-classifying galaxies based on kinematic data is not new; for instance, the taxonomy of the Hubble tuning fork can be reorganised on the basis of rotational properties of early-type galaxies (\citealt{cap2011}). }  Nevertheless, the fundamental physical processes which regulate galactic morphologies and kinematics remain largely uncertain.

We address this uncertainty by using numerical simulations to study the connection between the formation histories of S0 galaxies and their resulting morphologies and kinematics.  While all S0s share broadly similar properties { \bfd (e.g. dominant bulges and discs which lack spiral features)}, a growing body of literature considers S0s to be a grab-bag category encompassing a range of distinct formation histories (e.g. \citealt{lauri2010,barway2013,fm2018}). { \bfd This motivates a new approach which can clarify the fundamental physical mechanisms at play.}

In this work, we simulate a variety of { \bfd S0 formation pathways} including: isolated formation via disc instabilities (e.g. \citealt{noguchi98,saha2018}), tidal formation in a group halo (e.g. \citealt{byrd90,bekkicouch2011}), and formation via mergers (e.g. \citealt{bekki98,prieto2013}).  {\bfd To be clear, we do not consider S0 formation in dense environments, such as ram pressure stripping in galaxy clusters (e.g. \citealt{quilis2000}), which we leave to future work.  Nevertheless, we do not expect the ram pressure mechanism to differ significantly from other S0 formation mechanisms that we consider here.  For instance, the ram pressure scenario lacks violent global heating, which is also true for isolated disc instability driven by the local heating of clumps.

Our three formation pathways prevail in low-density environments such as groups and the field, and in each case we model the progenitor as a Spiral galaxy.  These pathways may be distinguished somewhat by the typical ratio of rotation to dispersion $v/\sigma$ in the S0 disc, with values ranging from $\sim 1-4$ for the isolated case, $\sim 0-1$ for the tidal pathway, and $\sim 1$ for mergers (e.g. see comparisons in \citealt{diaz2018}).

The goal of the present work is to train CNNs to extract fossil information from images and thereby predict various quantities associated with the formation of each S0.
To accomplish this, we first build a synthetic dataset consisting of morphological images of stellar density and spatial maps of line-of-sight velocities for simulated S0s. Our synthetic dataset is constructed to resemble observational images and kinematic maps\footnote{Throughout this paper we use the term `kinematic map' to refer strictly to the map of velocities projected along the line-of-sight to the observer, revealing the rotational pattern in the S0.  We use the term `morphological image' to refer to images of the stellar surface density.} of S0s, except with the added benefit that our simulations provide us with the exact formation history associated with each image.  We leverage this information from our simulations and train our CNNs to predict the correct formation pathway for each S0.  We also train CNNs to predict the merger mass ratio for those S0s which formed via mergers.}

The structure of this paper is as follows.  In Section \ref{sec:sim} we describe our overall set of S0 simulations including the parameter space that we explore.  In Section \ref{sec:syntheticdata} we describe how our simulations are transformed into synthetic datasets of morphological and kinematic images which comprise the training data for our CNNs.  In Section \ref{sec:nn} we outline the architecture of our CNNs, and we provide details on how we train the CNNs.  Section \ref{sec:res} describes the main results of the present work.  We provide a discussion in Section \ref{sec:discussion} including our thoughts on extending the present results to future analyses of observational data.  In Section \ref{sec:summary} we summarise and conclude.


\section{Description of N-body simulations} \label{sec:sim}

{ \bfd Our simulations model the transformation of Spiral galaxies into S0s using various physical mechanisms: disc instabilities, the tidal field of a group halo, and mergers.  To distinguish between these formation mechanisms } using CNNs, we must create a large synthetic dataset from N-body simulations comprising density maps and 2D kinematic maps.  {\bfd Our simulations are parameterised by numerous quantities which control the evolution within each scenario.}  It is important to ensure that our simulations are sufficiently diverse within the parameter space of possible interactions.  This will help to guarantee that our synthetic data is representative of each S0 formation path, and it will also help to prevent the CNNs from over-fitting to only a handful of examples with specific parameter values.

\subsection{Initial conditions} \label{sec:ic}

For each of the S0 {\bfd formation pathways} that we consider in this study, we assume that the progenitor is a Spiral galaxy.  {\bfd We construct the dark matter halo and the stellar disc to be similar to those of the Milky Way with parameterisations that are typical for models of Spiral galaxies (e.g. \citealt{bekki2015}).} For its dark matter halo, we choose a mass distribution following the NFW profile (\citealt{navarro96}) with a total mass of $10^{12} \Msun$ and a virial radius of 245 kpc.  {\bfd We choose a concentration of 10 based on correlations with halo mass in cosmological simulations (e.g. \citealt{neto2007}) }.  The stellar disc follows an exponential profile with a total mass of $6 \times 10^{10} \Msun$, a radial scale length of 3.5 kpc, a truncation radius of 17.5 kpc, and a gas mass fraction of $10\%$.

The main free parameter that we consider in the present work is the bulge-to-disc mass ratio (B/D).  Choosing a value for B/D determines the mass of the bulge $M_{\rm b}$ as some fraction of the mass of the disc.  This in turn determines the { \bfd half-light radius } of the bulge $R_{\rm b}$ through the Kormendy relation between stellar mass and radius (\citealt{kormendy77}).  The mass distribution of the bulge is specified by a Hernquist profile parameterised by $M_{\rm b}$ and $R_{\rm b}$.  For the present investigation, we construct three distinct models to use as initial conditions for our N-body simulations, with B/D and $R_{\rm b}$ values given in Table \ref{tab:ic}.  In each case, the truncation radius of the bulge is set to be five times the scale length.

The Toomre $Q$ parameter which controls the stability of the disc is set to a nominal value of 1.5 for the merger and group tidal simulations.  This ensures that any significant evolution of the disc is determined by external interactions.  For the isolated case, however, we set up an unstable disc with $Q=0$ by reducing the radial velocity dispersion to zero.  As a consequence, the disc can evolve significantly through the formation and eventual dissolution of clumps.  This choice is motivated in particular by the fact that unstable discs can evolve into S0-like remnants through isolated dynamical evolution alone (\citealt{saha2018,noguchi98}).

\begin{table}
  \begin{center}
  \caption{ {\bfd Quantities which distinguish our three progenitor Spiral models A, B, and C: bulge-to-disc mass ratios (B/D), half-light radius of the bulge ($R_{\rm b}$), and radial scale length of the disc ($R_{\rm d}$).  Approximate Hubble Type is also listed, which is estimated from the B/D ratios tabulated in Table 4 of \citet{gw2008}.}  Other parameters for these models are described in the text. }

  \begin{tabular}{lcccc}
  && Model A & Model B & Model C  \\
  \hline

  B/D & $-$ & 0.17 & 0.5 & 1.0 \\
  $R_{\rm b}$ &(kpc)& 3.5 & 6.1 & 8.6 \\
  $R_{\rm d}$ &(kpc)& 3.5 & 3.5 & 3.5 \\
  Hubble Type & $-$ & Sb/Sbc & Sa & S0/a \\
  
  \hline

  \label{tab:ic}
  \end{tabular}
  \end{center}
\end{table}

\begin{table}
  \begin{center}
  \caption{Summary of the range of parameter values from which random values are drawn for our tidal and merger scenarios. Full descriptions of the parameters are given in the text.}
  \begin{tabular}{clcc}
 
  \multicolumn{4}{c}{ {\bfd Tidal} } \\
  \hline
  Parameter & Description & unit & Range of values \\
  \hline
  $M_{\rm halo}$ & Total group mass & ($10^{13}~\Msun$) & $1.0 - 6.0$ \\
  $r_0$ & Initial position & ($R_s$ of halo) &  $1.0-3.0$ \\
  $v_0$ & Initial velocity & ($V_{\rm cir}$ at $r_0$) & $0.2 - 0.8$ \\
  $\theta$ & Polar angle & ($\deg$) & $0 - 180$ \\
  $\phi$ & Azimuthal angle & ($\deg$) & $0 - 360$ \\
  
  & & & \\
  
  \hline
  \multicolumn{4}{c}{Sp-Sp Mergers} \\
  \hline
  Parameter & Description & unit & Range of values \\
  \hline
  $m$ & Mass ratio & - & $0.05-0.4$ \\
  $r_{\rm peri}$ & Pericentre & (kpc) & $17-50$ \\
  $e_{\rm orbit}$ & Orbital eccentricity & - & $0.4-0.9$ \\
  $\theta_{\rm A}$ & Polar angle A & ($\deg$) & $0 - 180$ \\
  $\phi_{\rm A}$ & Azimuthal angle A & ($\deg$) & $0 - 360$ \\
  $\theta_{\rm B}$ & Polar angle B & ($\deg$) & $0 - 180$ \\
  $\phi_{\rm B}$ & Azimuthal angle B & ($\deg$) & $0 - 360$ \\
  \hline

  \label{tab:par}
  \end{tabular}
  \end{center}
\end{table}

\subsection{Parameters for the tidal simulations}

For the tidal interaction scenario, we place a given Spiral model in an orbit around a fixed gravitational potential representing a group-scale halo.  {\bfr This treatment does not consider galaxy-galaxy interactions within the group, only the tidal interaction with the smooth gravitational potential of the group itself.}  We consider spherical group halos given by the NFW profile with a total mass in the range $1.0-6.0 \times 10^{13} ~ \Msun$.  The concentration parameter for each group halo is computed as a function of its mass in accordance with the results of cosmological simulations (e.g. \citealt{neto2007}). We place the Spiral galaxy at a distance $r_0$ from the centre of the group halo, where $r_0$ is considered to be some multiple of the NFW group halo's scale radius $R_s$ as given in Table \ref{tab:par}.

The initial velocity $v_0$ of the galaxy is oriented in a perpendicular direction to its position vector, and its magnitude is considered to be some fraction of the velocity $V_{\rm cir}$ needed to maintain a circular orbit at $r_0$.  This choice initialises the orbit at its apocentre, which allows our initial equilibrium model to gradually evolve under tidal forces as it falls into the group halo for the first time.  The orientation of the disc is given by the polar angle $\theta$ and azimuthal angle $\phi$, with values varying over the range given in Table \ref{tab:par}.

\subsection{Parameters for the merger simulations}

For the merger scenario, we place two Spiral models in an eccentric mutual orbit and rescale one of the models for a given mass ratio.  We choose eccentric orbits $e_{\rm orbit}$ which are likely to lead to mergers on the fixed timescale of our simulations, as given in Table \ref{tab:par}.  We explore mass ratios $m$ in the range $0.05-0.4$ because smaller values will not yield mergers over the timescales we consider, and larger values can yield violent mergers which may destroy the disc and therefore fail to produce S0-like remnants.

We separate the galaxies by an initial distance $r_0$ corresponding to the mean of the pericentre and apocentre, which we calculate from the chosen values of $r_{\rm peri}$ and $e_{\rm orbit}$.  We truncate the maximum value of $r_0$ at 170 kpc so that highly eccentric orbits (i.e. those with very large apocentres) may have time to merge within the time window of the simulation.

The orientation of the primary Spiral is given by the polar angle $\theta_{\rm A}$ and azimuthal angle $\phi_{\rm A}$, and the corresponding angles for the secondary Spiral are  $\theta_{\rm B}$ and $\phi_{\rm B}$.  As with the tidal scenario, values for these angles are drawn from a uniform distribution over their full range as indicated in Table \ref{tab:par}.

\subsection{Dynamical evolution}

We adopt the GPU-accelerated numerical code described in detail in our previous work (e.g. \citealt{bekki2013,bekki2014}).  { \bfr Whereas the gravitational dynamics are computed on GPUs, all other calculations are performed on the CPU, including gas dynamics and star formation.  Further details on the simulation code are presented in Appendix \ref{sec:code}.}

We ran each simulation on a server equipped with a GeForce GTX 1080 Ti at the University of Western Australia.  For a given initial Spiral model (A, B, or C in Table \ref{tab:ic}), we choose 100 random combinations of parameter values for the tidal scenario and 100 random parameter sets for the merger case from the ranges in Tables \ref{tab:par}.  For the isolated scenario, we simply run the $Q=0$ version of each of the initial Spiral models.  This yields a total of 603 simulations.  In each case, we evolve our models for a total of 5.6 Gyr with a fixed timestep of 1.4 Myr.

\section{Description of synthetic data} \label{sec:syntheticdata}

Here we describe how we convert the outputs of our simulations into input data for our CNNs.  First we must judge which simulations result in the formation of S0s and at what times.  Then we describe our process for creating images of the morphology and kinematics of the selected galaxies.

\subsection{Criteria for selecting S0s} \label{sec:sel}

\subsubsection{Isolated Models}

For our isolated pathway of S0 formation, there are only three total simulations to consider: the Q=0 version of the initial disc for the initial models A, B, and C (see Table \ref{tab:ic}).  When simulating each of these models in isolation, the disc rapidly forms clumps and other substructures owing to its inherent instability.  The clumps coalesce into the centre of the galaxy over time { \bfd which builds up the bulge, resulting in final B/D ratios of 0.47, 0.78, and 1.22 for models A, B, and C, respectively\footnote{The final B/D values for the isolated models are determined by a two-component Sersic fit to the one-dimensional surface density profile in the plane of the disc.}.  Given initial B/D values of 0.17, 0.5, and 1.0 (see Table \ref{tab:ic}), this means the discs lose 20\%, 16\%, and 10\% of their initial mass, respectively, due to the formation and migration of clumps.

\citet{saha2018} explore the same mechanism of dynamical instability in the context of S0 formation and focus on the evolution of an initial disc with a negligible bulge (${\rm B/D} \approx 0.03$).  In their simulations, the formation and migration of clumps leads to the formation of a final S0 with ${\rm B/D} \approx 0.6$.  This corresponds to a mass loss of 35\% for the disc as it builds the bulge, which is somewhat higher than the values for our simulations.}

{\bfd Following the phase of bulge build-up, the discs of our isolated models are relatively featureless apart from a central bar.}  We consider the {\bfd initial Spiral} to have transformed to S0 under this scenario when substructure {\bfd (e.g. clumps)} are no longer dominant features of the morphology.  This occurs in all cases around ~1.5 Gyr after the start of the simulation.

\subsubsection{Tidal Models}

Not all of the simulations will form S0s.  For the tidal models, numerous models never pass close enough to the centre of the group halo to undergo tidal processing.  This is a result of having large values for both $r_0$ and $v_0$ as given in Table \ref{tab:par}.  In contrast, other models pass too close to the group centre, suffering significant disc disruption.  For most models, there are one or more close encounters with the group centre which heat the disc and might also disrupt its outskirts.

{\bfd Without knowing a-priori the ideal parameters to produce S0s, we opt to verify the transition from Spiral to S0 by visual inspection for each of the simulations.  Primarily we inspect the two-dimensional surface density for a coherent disc without spiral arms as well as a visually dominant bulge.  This allows us to exclude simulations in which tidal forces or mergers are too strong and destroy the disc, and it also excludes simulations where the interactions are weak and the progenitor remains a Spiral galaxy.  Because we wish to create a diverse set of simulations for training our CNNs, we choose not to apply any restrictions on other properties such as kinematics, presence of bars, star formation, etc.

After cross-checking to see which parameter combinations and orbits yield a good collection of simulated S0s,} we apply cutoffs on orbital properties as follows.

First, we require that at least one pericentre in the orbit passes within 130 kpc of the group centre.  If a pericentre is 30 kpc or less, we exclude all subsequent evolution from consideration.  Of the remaining orbits, we consider the first pericentre to be the initial phase of transformation into S0.  However, the close encounters will in general remove some material from the disc (e.g. producing tidal tails), such that the galaxy would be categorised as irregular.  We find that it takes ~$0.5-0.75$ Gyr for these irregularities to resettle to equilibrium and visually dissipate. {\bfd We primarily consider the Spiral to have transformed to S0 if after the tidal interaction the galaxy contains a disc without spiral features.  We classify the galaxy as S0 from this point in time up until the next pericentre; we then iteratively apply these same criteria at each pericentre until the end of the simulation is reached. }

\begin{figure}
   \begin{center}
   \includegraphics[width=0.45\textwidth]{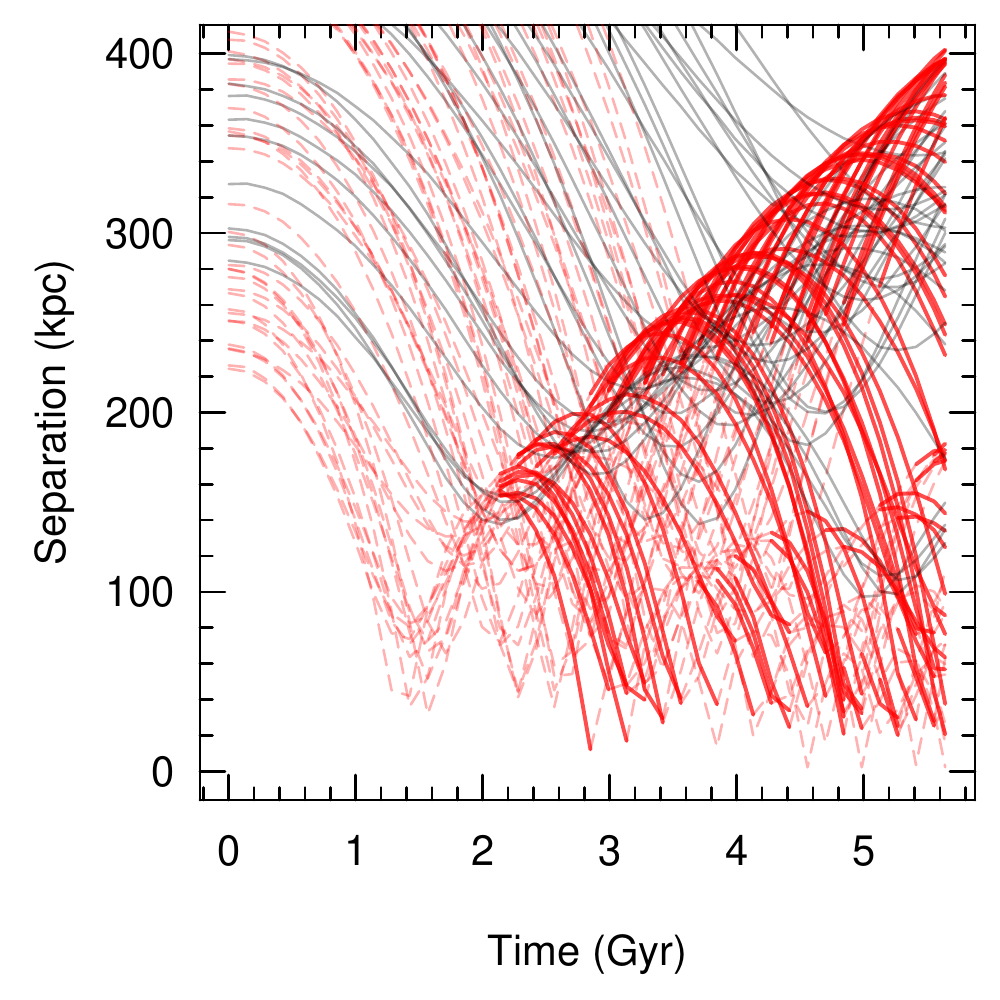}
   \caption{Orbits for the tidal interaction scenario, {\bfd showing a trace of radius versus time for 100 unique parameter combinations } (see Table \ref{tab:par}).  The separation between the simulated galaxy and the centre of the group halo is shown from T=0 (the start of the simulation) to T=5.6 Gyr (the end).  Simulations which do not pass our criteria for being considered an S0 as described in Section \ref{sec:sel} are shown in grey.  For the remaining simulations, positions in the orbit which pass our criteria are shown as solid red lines, and all other orbital positions are shown as dashed red lines.}
   \label{fig:orbits_tidal}
   \end{center}
\end{figure}

Imposing these criteria on our full set of {\bfd group simulations, we find that 64\% of Spirals} undergo a transformation to S0.  Figure \ref{fig:orbits_tidal} shows the unique orbits for our set of tidal simulations, where each orbit is characterised by a random combination of parameter values from the ranges given in Table \ref{tab:par}. Red lines are those which pass our criteria at some point in their evolution, with solid and dashed linestyles distinguishing the timesteps which do and do not pass the criteria, respectively.  Grey lines are those orbits which never pass our criteria.

\subsubsection{Merger Models}

For the merger case, a number of models do not merge within the time frame (5.64 Gyr) of the simulation, particularly for small mass companions (e.g. $ m \approx 0.05 $).  The force which drives the merger (dynamical friction) scales as the squared mass of the satellite, which means that the timescale of the merger is highly dependent on the satellite mass (e.g. \citealt{BT}).  We find that the minimum mass ratio required to create a merger in our fixed time frame is $\approx$ 0.1.

As in the tidal scenario, we consider S0s to have formed in the merger scenario if the morphology of the merger remnant {\bfd contains a bulge and disc without spiral features}.  We visually inspect the simulations to see which mergers yield such results and at what times.  We capture these good models into our final set of S0s by imposing the following criteria: the separation between the centres of the two galaxies must remain below 10 kpc for a duration of at least 0.75 Gyr.  For convenience, we estimate each centre as the position of the particle at the true centres-of-mass at T=0.  When tracked in this manner, the separation between the centres will not necessarily tend toward zero in our mergers.  For the smaller galaxy in particular, this central particle may be displaced from the true centre owing to disruption into streams, rings, or other substructure within the disc.

\begin{figure}
   \begin{center}
   \includegraphics[width=0.45\textwidth]{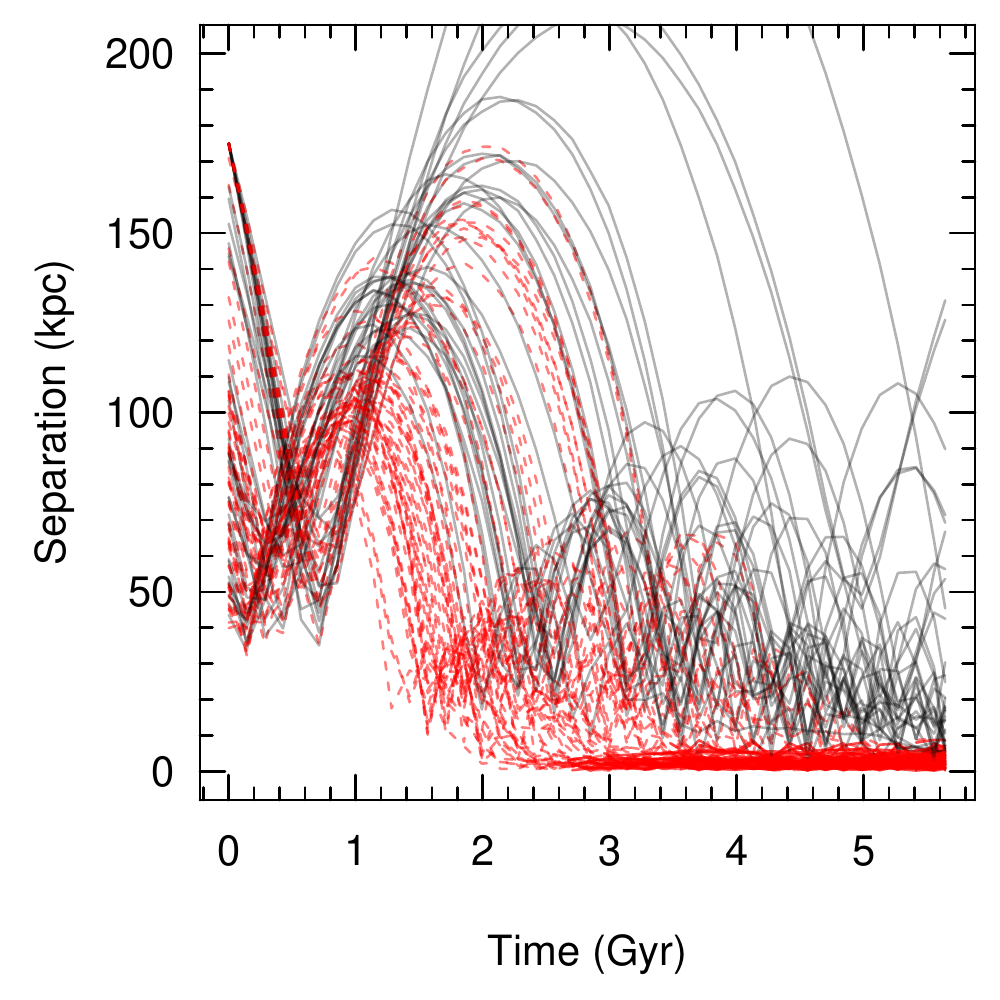}
   \caption{Orbits for the merger scenario, {\bfd showing a trace of radius versus time for 100 unique parameter combinations } (see Table \ref{tab:par}).  The separation between the two galaxies is shown from T=0 (the start of the simulation) to T=5.6 Gyr (the end).  Orbits which passed our criteria for merging into an S0 as described in Section \ref{sec:sel} are shown in red, with a dashed linestyle prior to the merger, and solid linestyle after the merger.  Those simulations which did not pass the merger criteria are shown in grey.}
   \label{fig:orbits_merger}
   \end{center}
\end{figure}

Orbits which do not yield a merger are shown as grey lines in Figure \ref{fig:orbits_merger}.  As stated above, mass ratios of $\approx$ 0.1 and less are those which do not create mergers.  The other orbits (red in Figure \ref{fig:orbits_merger}) span the range of mass ratios $0.1-0.4$.  Overall, we find that none of the mergers are strong enough to significantly disrupt the disc of the primary, due to the fact that we do not explore mass ratios larger than 0.4.  We check visually that the morphologies of the merger remnants do indeed resemble S0s, and those which do not pass this step of visual inspection are removed from the sample.  Under these criteria, 65\% of our {\bfd Sp-Sp} merger simulations produce an S0.

\subsection{Creating input images for the CNNs} \label{sec:resref}

Our synthetic data is compiled from the models which produce S0s as described in the previous section.  This gives us a total of 64 tidal models, 65 merger models, and one isolated model for each initial condition A, B, and C (Table \ref{tab:ic}), which sums to 390 total N-body models.  {\bfd For a given model, the galaxy is considered an S0 at a specific range of times, shown graphically as the solid red lines in Figures \ref{fig:orbits_tidal} and \ref{fig:orbits_merger}.  Within those range of times, we take snapshots of the model at time intervals of 140 Myr.  We then produce images of mass surface density and mass-weighted velocities projected at 50 random orientations, {\bfr varying both the inclination of the disc as well as the azimuthal angle within the disc plane.}}

Summing across all valid timesteps for the models, our synthetic dataset has a total of 266,550 images of stellar density, and the same number of two-dimensional kinematic images.  The respective total for each S0 {\bfd formation pathway} is 131,100 images for mergers, 130,650 for tidal, and 4,800 for isolated.

{\bfd The focus of the present work is to validate our new methodology on simulated data, but our choice of image parameters such as physical scale and resolution is motivated by recent observational datasets.}  For instance, the SAMI Galaxy Survey obtains spatially resolved spectra for many thousands of galaxies with spatial resolutions {\bfd of $0.5~\arcsec$ across a diameter of 31 pixels.  This translates to physical sizes of $1.2-26.6$ kpc for each galaxy at resolutions of $0.1-2.8$ kpc per pixel} (\citealt{bryant2015}).

{\bfd We choose nominal values within these ranges for image scale and resolution, placing each S0 at the centre of a $20 \times 20$ pixel image with a fixed size of $18~\kpc$ per side, giving a resolution of 0.9 kpc per pixel.  These fixed values suffice for our present purposes, but in the future we will strive to create simulated datasets which reproduce the observations accurately.  To accomplish this, we would need to use a range of scales and resolutions for our images as well as consider other factors such as noise.  While we have not added any noise to our images, it will be important to do so in the future to mimic observational conditions.}

{\bfd To create our morphological images, we compute the surface mass density of the stellar particles} on a logarithmic scale in the range $10^6 - 10^{10} ~\Msun \kpc^{-2}$.  {\bfd To create our kinematic maps, we compute the mass-weighted average velocity of all stellar particles in each spatial bin in the range $-200$ to $200 ~\kms$.  We perform this computation in the rest frame of each S0, and we consider only the line-of-sight velocity for each given projection.  For both morphological images and kinematic maps}, the value of any bin exceeding the lower or upper limit is set to the respective bounding value.

If a spatial bin does not contain any particles, we must set its pixel value by hand.  We must do so because missing values would be passed into a neural network as { \bfd non-numerical values or infinities } which would then propagate into the weights of the network and break its training.  For the morphological images, pixels with no data are assigned to the value of the lower bound.  This corresponds naturally to a black sky background.

{\bfd For the kinematic images, however, setting the value of empty pixels to the lower ( upper) bound improperly assigns that pixel a large negative (positive) velocity.  We therefore assign empty pixels a velocity of $0~\kms$ to match the rest frame of each galaxy, which corresponds to the midpoint in the range of possible pixel values.}

{ \bfd Nearly all images have empty pixels, apart from several images of face-on S0s.  The fraction of empty pixels can be large for some images, e.g. up to $\sim 75\%$ for edge-on systems.  On average across the full set of images, the fraction of empty pixels in each image is only $\sim11\%$.}

\begin{figure*}
   \begin{center}
   \includegraphics[width=1.0\textwidth]{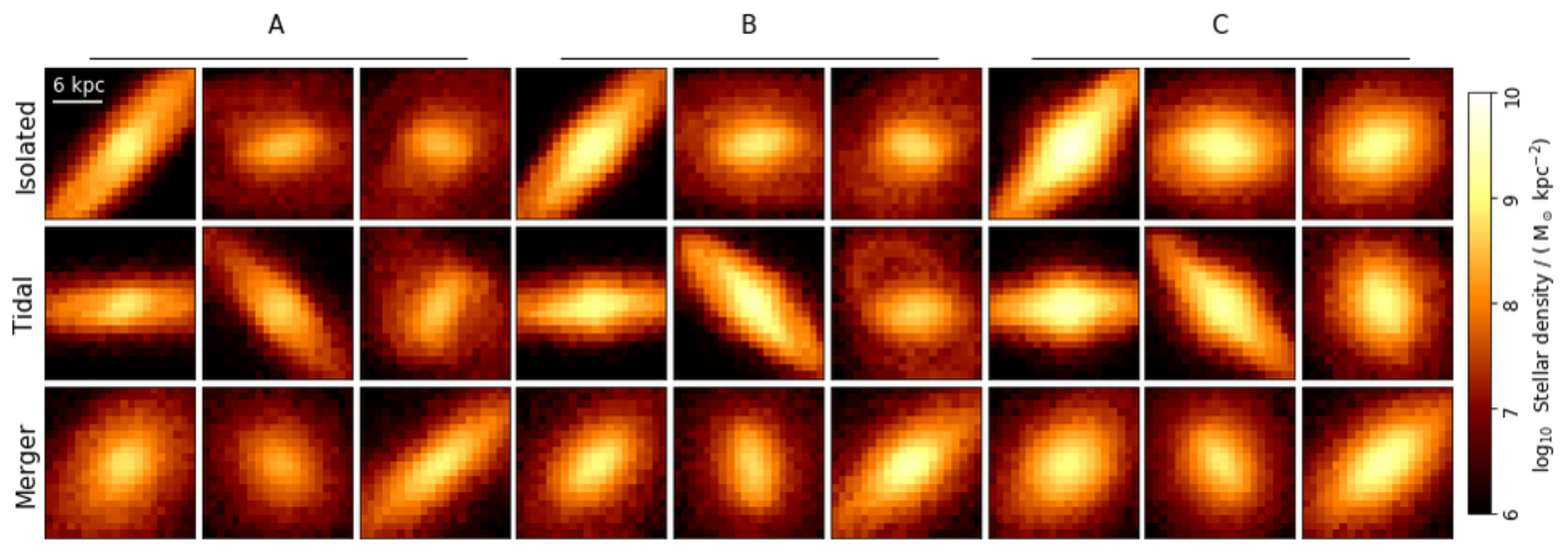}
   \caption{ {\bfd Images of S0s drawn randomly from our synthetic dataset, shown as the logarithm of the stellar surface mass density.  Colour is used for visual illustration only.  The inset in the upper left panel shows the physical scale for each image (the field of view is 18 kpc $\times$ 18 kpc for all images).}  See text for further details on the dataset.  Each row pertains to a different {\bfd formation pathway}: isolated (top), tidal (middle), and merger (bottom).  The first three columns correspond to the Spiral progenitor model A, the middle three columns to model B, and the final three columns to model C (see Table \ref{tab:ic}).  Despite the different formation paths and initial conditions, the morphology of these simulated systems appear to be quite similar, much like observed S0s.  This similarity underscores the difficulty of the classification task.}
   \label{fig:den}
   \end{center}
\end{figure*}

\begin{figure*}
   \begin{center}
   \includegraphics[width=1.0\textwidth]{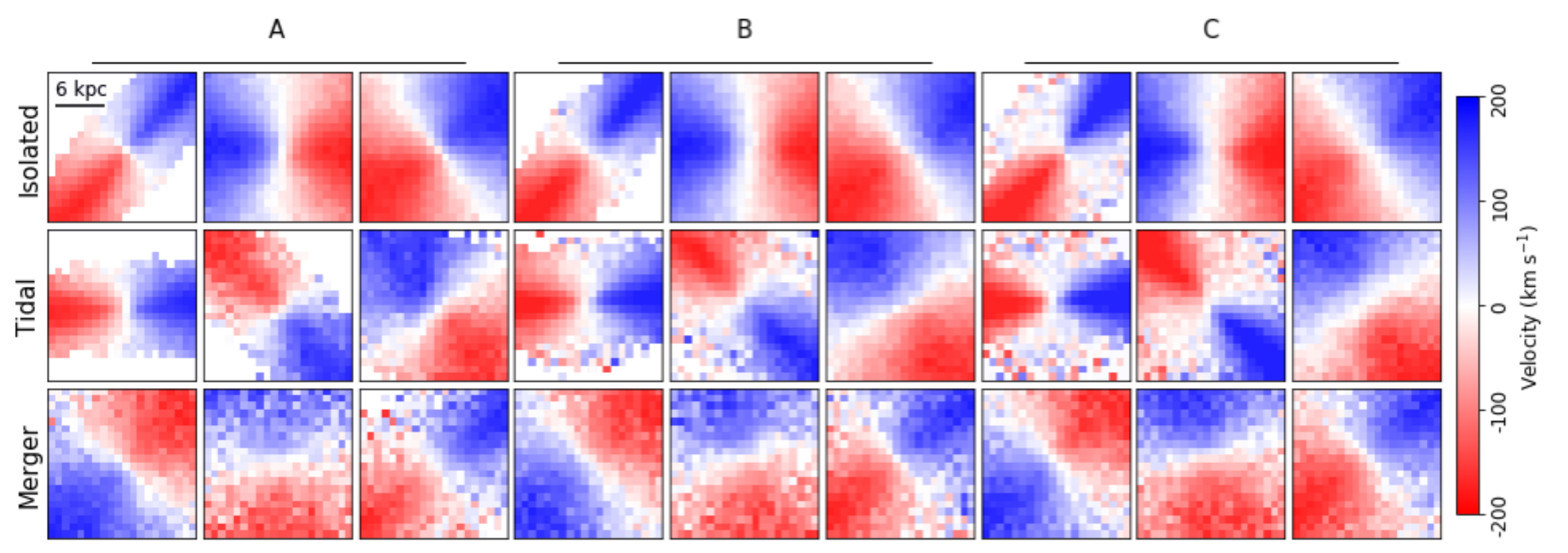}
   \caption{Two dimensional maps of the mass-weighted average velocity corresponding to the galaxies shown in Figure \ref{fig:den}.  {\bfr Colour is used for visual illustration only.}  {\bfd Physical scale, formation pathway, and progenitor Spiral model are the same as described in Figure \ref{fig:den}. }}
   \label{fig:vel}
   \end{center}
\end{figure*}

Figures \ref{fig:den} and \ref{fig:vel} show morphological and kinematic images, respectively, drawn randomly from our synthetic dataset.  Despite differences in {\bfd formation pathway} and initial conditions, many of these images appear quite similar, suggesting that visual classification would be challenging and time-consuming for humans.  This underscores the scientific role that CNNs could potentially fulfil by supplementing the abilities of astronomers.

{\bfd We emphasise that the images in Figures \ref{fig:den} and \ref{fig:vel} are shown in colour for visual illustration only.  When passed into the CNNs, the images are monochromatic.  We also emphasise that the presence of randomness within some images, particularly in Figure \ref{fig:vel}, is intrinsic to the simulations (e.g. dispersion of the bulge or heating of the disc).  No artificial noise was added to the images, as stated previously. {\bfr Several high-resolution images of representative S0 simulations are shown in Appendix \ref{sec:hires} for comparison. } }

\subsection{Preparing the training data} \label{sec:preptrain}

\begin{table*}
  \begin{center}
  \caption{Description of our trained CNNs, comprising the main results of this work.}

  \begin{tabular}{lcccc}
  \hline
  & Scientific prediction & Task & Input image type & Input image shape  \\
  \hline
  Model 1a & S0 {\bfd formation pathway}  & Classification  & Morphology \& Kinematics & $20\times20\times2$ \\
  Model 1b & "  & "  & Morphology & $20\times20$ \\
  Model 1c & "  & "  & Kinematics & $20\times20$ \\
  \hline
  Model 2a & Merger mass ratio  & Regression & Morphology \& Kinematics & $20\times20\times2$ \\
  Model 2b & "  & "  & Morphology & $20\times20$ \\
  Model 2c & "  & "  & Kinematics & $20\times20$ \\
  
  \label{tab:nn}
  \end{tabular}
  \end{center}
\end{table*}

In the present work, we have two scientific tasks: to train a CNN to classify S0 {\bfd formation pathways}, and to train another CNN to predict merger mass ratios. {\bfd In principle, we can create a decision tree whereby our first CNN predicts which galaxies are formed by mergers and passes them to the second CNN which then estimates the merger mass ratios.  We do not explore this approach in the present work, however.  Our two CNNs are independent of one another.}

For each {\bfd of these scientific tasks}, we perform three experiments.  We train one network on the morphological images only, we train a second network on the kinematic images only, and we train a third network on both the morphological and { \bfd kinematic maps}.  Table \ref{tab:nn} provides a summary of the CNNs that we train.

In the third case (which we will take to be our main results), matching pairs of images are passed into the network as separate channels of an image array.  That is, an S0 at a given time and at a given orientation will be represented by an image array of size $20 \times 20 \times 2$, with the morphological and kinematic images occupying different slices in the final dimension.  When passing the image data into the network, pixel values in the range 0 to 255 are rescaled to the range 0.0 to 1.0 as is typical for machine learning tasks.

Rather than using our full synthetic dataset to train the networks, we take a {\bfd random 80\% fraction} of the dataset as our training data.  The remainder of the full dataset is known as the test set, which will be used to validate the predictions of our trained network. In other words, it is important to verify that the network provides accurate predictions for both its training data as well as new data which it has not been exposed to.

\subsubsection{Classifying S0 formation pathway}

For our first scientific task, we must train a network to predict formation paths from a set of input images.  We take a random 80\% and 20\% sampling of the full dataset to create our training and test data, respectively.  This gives us 213,240 images in the training set, and 53,310 in the test set.

As a supervised learning task, we must assign a categorical label to each input image.  These labels identify the {\bfd formation pathway} as `Isolated', `Tidal', or `Merger', which we convert to the numerical labels 0, 1, and 2, respectively.

\subsubsection{Predicting merger mass ratio} \label{sec:mehmrat}

\begin{figure}
   \begin{center}
   \includegraphics[width=0.5\textwidth]{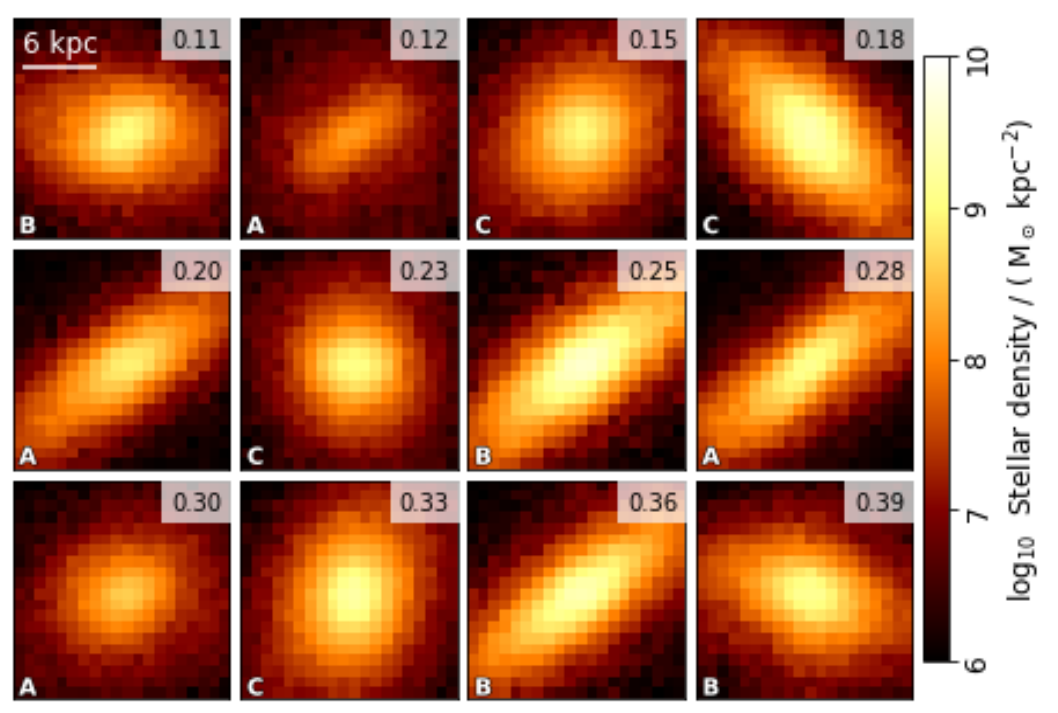}
   \caption{{\bfd Images of S0s formed via mergers drawn randomly from our synthetic dataset, shown as the logarithm of the stellar surface mass density. \bfd Colour is used for visual illustration only. The merger mass ratio is written in the top right corner of each panel, ranging from 0.11 (top left panel) to 0.39 (bottom right panel).  The progenitor Spiral model associated with the primary galaxy is noted in the bottom left corner of each panel (A, B, or C; see Table \ref{tab:ic}).  Also drawn in the upper left panel is the physical scale for each image (the field of view is 18 kpc $\times$ 18 kpc for all images). } }
   \label{fig:merger_den}
   \end{center}
\end{figure}

\begin{figure}
   \begin{center}
   \includegraphics[width=0.5\textwidth]{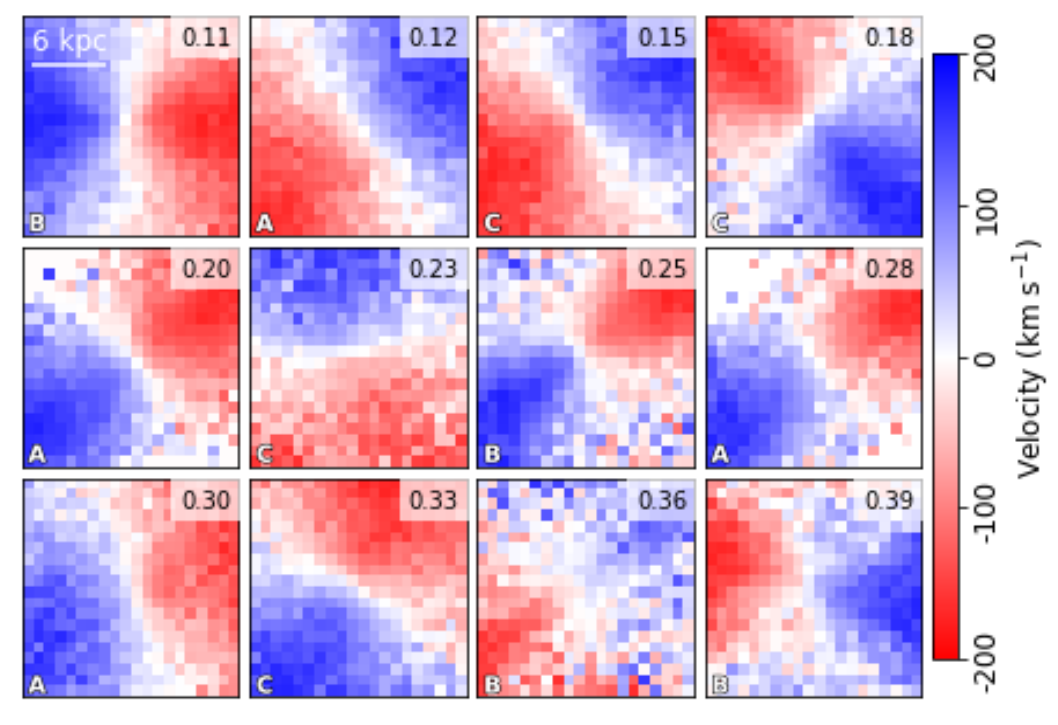}
   \caption{Two dimensional maps of the mass-weighted average velocity corresponding to the galaxies shown in Figure \ref{fig:merger_den}.  {\bfr Colour is used for visual illustration only.} {\bfd Physical scale, mass ratio, formation pathway, and progenitor Spiral model are the same as described in Figure \ref{fig:merger_den}. }}
   \label{fig:merger_vel}
   \end{center}
\end{figure}

For the task of predicting mass ratios, we discard all input images except those which pertain to the mergers.  To each of these images we assign a numerical label equal to the mass ratio which was used as an initial parameter in the N-body simulation (Table \ref{tab:par}).  In Figures \ref{fig:merger_den} and \ref{fig:merger_vel} we show a random sample of these morphological and kinematic images, respectively, {\bfd and we also label the associated mass ratio and progenitor Spiral model in each panel.  As in Figures \ref{fig:den} and \ref{fig:vel}, the colours shown in Figures \ref{fig:merger_den} and \ref{fig:merger_vel} are for visual illustration only.  When passed into our CNNs, the images are monochromatic.}

\begin{figure}
   \begin{center}
   \includegraphics[width=0.45\textwidth]{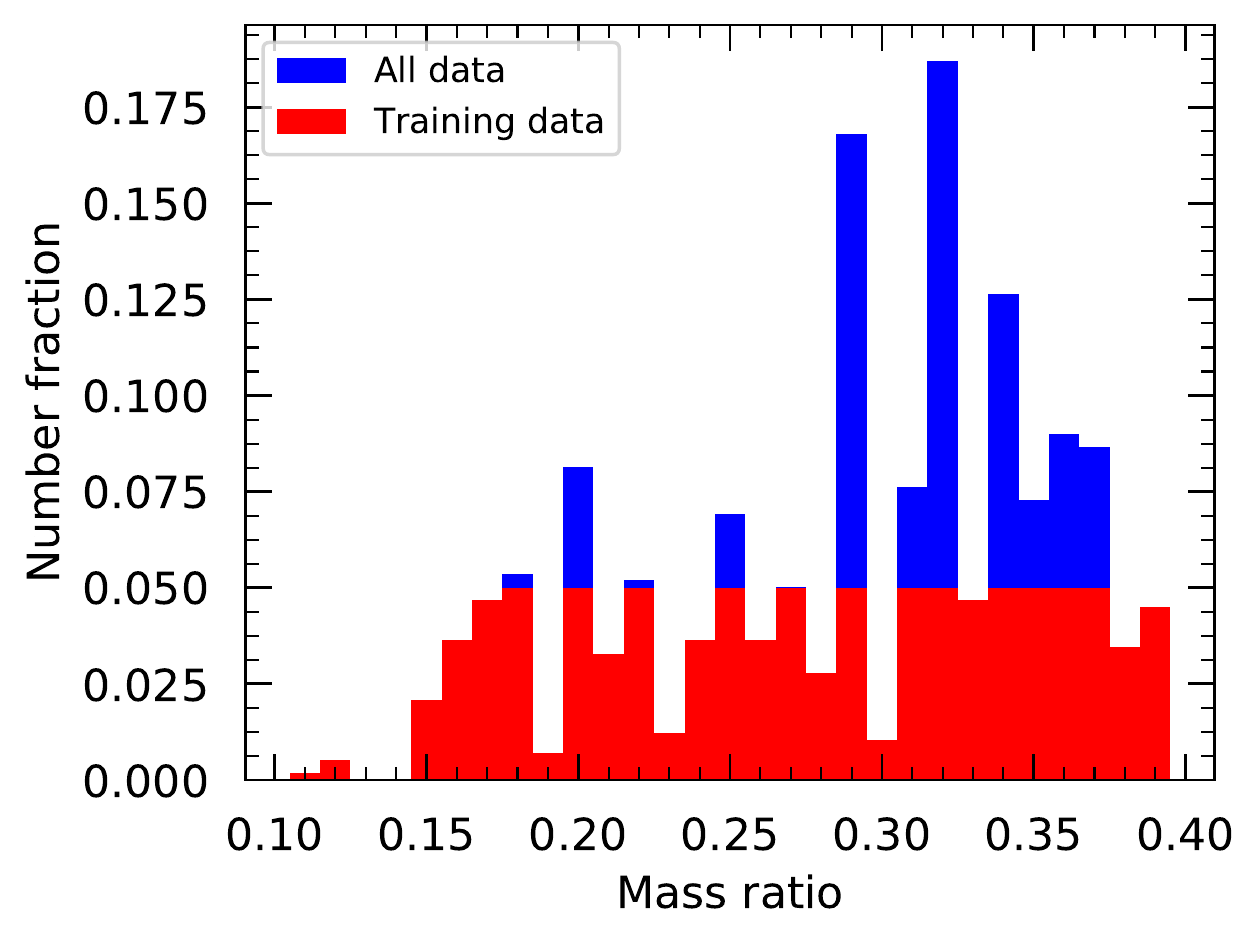}
   \caption{Histogram of mass ratios for the S0s formed via mergers in our synthetic dataset.  The heights of the red bars represent the relative fraction of each mass ratio in the training data.  By construction, this fraction does not exceed 5\% for any given mass ratio in the training data (see Section \ref{sec:mehmrat} for details).  Meanwhile, the heights of the blue bars correspond to the distribution of mass ratios in the full dataset of S0 mergers.}
   \label{fig:merger_hist}
   \end{center}
\end{figure}

To split our data into training and test sets, we do not take random samples as we did before.  This is because a random sampling of the merger data will be biased toward certain mass ratios which dominate the dataset.  Figure \ref{fig:merger_hist} shows the non-uniform distribution of this data as a function of mass ratio.  {\bfr Some of the non-uniformity is simply due to random sampling of simulation parameters, resulting in some mass ratios being represented by a larger pool of simulations as compared to others.

However, the peaks in the histogram generally indicate a true preference to form S0s by mergers for mass ratios in the range $0.28-0.34$.  Several factors may contribute to this, including artificial issues related to how we configure the simulations (e.g. small mass ratios requiring longer timescales to merge than we considered in our setup of the N-body simulations), as well as genuine physical reasons (e.g. larger mass ratios capable of disrupting the disc).} To ensure that our network is not significantly biased toward over-represented mass ratios during training, we need our training data to have a roughly uniform distribution across the full range of mass ratios.

To construct such a training set, we require that all data with a given mass ratio constitute at most 5\% of the overall dataset.  Figure \ref{fig:merger_hist} shows the distribution of this training set in red, which sums to a total of 86,622 images.  The remaining 44,478 images in the merger dataset are used as the test set, resulting in a fractional split between the training and test data of 66\% and 34\%, respectively.

\section{Description of Neural Networks} \label{sec:nn}

\subsection{Architecture}

As demonstrated by previous studies, neural networks composed of only a handful of convolutional layers can be trained to successfully classify galaxy images (e.g. \citealt{dieleman2015,ds2018}).  Meanwhile, complex state-of-the-art convolutional networks with a multitude of layers (i.e. `deep' networks) have also proven to be very successful at classifying galaxies, even though these architectures were originally devised for general purpose image classification (e.g. \citealt{dai2018, ackerman2018}).

\begin{figure*}
   \begin{center}
   \includegraphics[width=0.95\textwidth]{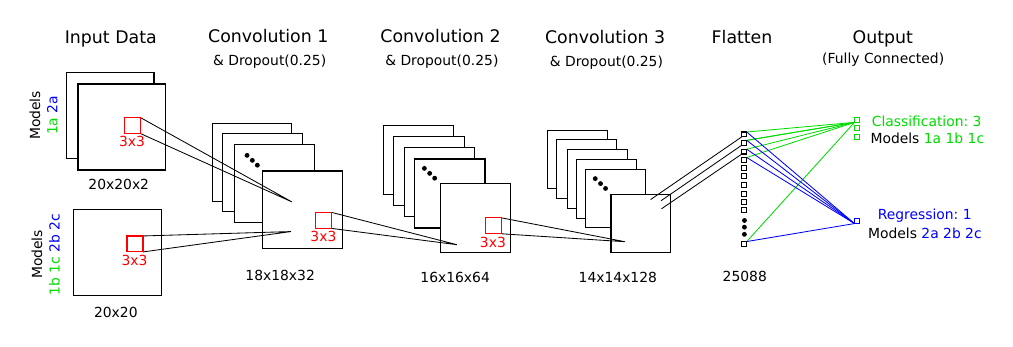}
   \caption{Schematic of the CNN architecture adopted in the present study, with further details provided in the text.  To train our six models (see Table \ref{tab:nn} for the model labels), we use the same basic network architecture displayed in this figure, with changes applying only to the final layer and the input data.  Also labelled are the dimensions of a single image (or image array) as it is fed into the network and is passed through each of the layers.  For the regression models, the output of the network is a single number: the predicted mass ratio.  For the classification models, the output consists of three numbers representing the predicted probabilities for each class (isolated, tidal, merger); we take the class with largest probability as the prediction of the network.}
   \label{fig:nn}
   \end{center}
\end{figure*}

In the present work, we adopt a simple network architecture rather than a deep network.  We are guided by the notion that our methods should be no more complex than required by our scientific goals.  Figure \ref{fig:nn} shows a schematic of the adopted network architecture, with variations labelled for each of our CNNs as given in Table \ref{tab:nn}.  In each case, we use only three convolutional layers, which is fewer than that of various previous studies which have adopted four convolutional layers (e.g. \citealt{dieleman2015,ds2018}).

The input layer of our CNNs depends on the physical data being passed into the network.  For models trained on both morphological images and kinematic maps (i.e. models 1a and 2a as given in Table \ref{tab:nn}), the input is an image array of size $20\times20\times2$, where the final dimension denotes the separate image channels assigned to the morphology and kinematics, respectively.  For models trained on either morphology or kinematics (i.e. models 1b, 1c, 2b, and 2c), the input image shape is simply $20\times20$.  These differences in the input layer are illustrated in the leftmost column of the schematic Figure \ref{fig:nn}.

The next layer in our architecture convolves a kernel of size $3\times3$ with the given input for each of 32 total convolutional filters.  A nonlinear activation known as `relu' (rectified linear unit) is then applied, which has the effect of setting any negative output values to zero.  Once the output of each of the 32 convolutions are stacked together, an array of size $18\times18\times32$ is produced.  We then follow this operation with a dropout layer, which randomly sets a given fraction of inputs to zero at each update during training time.  We choose the dropout fraction to be 0.25.  The effect of the dropout layer is to prevent the parameters of the network from being tuned to any one particular feature that is produced from the preceding layer.  In other words, dropout helps to prevent overfitting.

Following this, we have two additional convolutional layers along with dropout layers.  These additional layers have the same structure as before, except that the number of $3\times3$ filters in the second and third convolutional layers is increased to 64 and 128, respectively.  Consequently, the output of the second convolutional layer has size $16\times16\times64$, and the output from the third convolution has size $14\times14\times128$.  We then flatten this image array into a one-dimensional output vector of length 25,088.  The final layer of the network is the fully connected output which performs linear combinations of the 25,088 values from the previous layer.

The nature of the fully connected layer depends on whether the network is performing classification (green in the rightmost column of Figure \ref{fig:nn}) or regression (blue).  For classification networks (models 1a, 1b, 1c), the linear combination is performed at each of three independent nodes, one for each S0 {\bfd formation pathway} (isolated, tidal, merger).  A `softmax' activation is applied to the layer to guarantee that the output values for the three nodes sum to one.  We can then interpret these three values as the predicted probabilities that the given input image is a member of each of the respective classes.  The class with the largest probability is taken to be the predicted class for that input image.

For regression networks (models 2a, 2b, 2c), the fully connected layer contains a single node and no activation function is applied. We interpret this single number as the predicted mass ratio generated by the network for the given input image. 

We arrived at the adopted architecture by an ad-hoc process of stacking variations of the convolutional layers with variations of dropout layers and training each architecture on the same data.  By tweaking layers and their hyperparameters, we eventually found acceptable results with the adopted architecture.  It is possible that this architecture can be further simplified while retaining equivalent or perhaps marginally improved results, but doing so is beyond the scope of the present work.  We focus instead on satisfying our scientific goals as detailed in Section \ref{sec:res}, and we leave the exploration of an optimised or minimal network to future work.

\subsubsection{Training}

The trainable parameters in the convolutional layers are the pixel values within each of the kernels.  With $3\times3$ kernels applied to an array of $N$ input images plus one bias parameter, the total number of parameters for a given kernel is $3\times3\times N + 1$. For the first convolutional layer, the number of trainable parameters for each of the 32 kernels is either 10 (for models 1b, 1c, 2b, 2c) or 19 (for models 1a, 2a), which sums to 320 and 608 parameters in total, respectively.  Similarly, the number of trainable parameters in the second and third convolutional layers can be summed to 18,496 and 73,856, respectively.

In addition to the convolutional layers, trainable parameters also occur in the final fully connected layer as weights in the linear combinations.  For regression networks (models 2a, 2b, 2c), there is one bias parameter plus one weight per input value, which sums to 25,089 total parameters in the final layer. For classification networks (models 1a, 1b, 1c), the same number of parameters exist for each of three nodes, totalling to 75,267 parameters in the fully connected layer.

Summing across all layers of each network, the classification networks have $\approx168,000$ total trainable parameters, and the regression networks have $\approx118,000$ parameters.  These parameters are initialised to random values prior to training.

The goal of training each network is to adjust the values of its parameters so that its predictions on the training data are optimised against a given cost function. For our classification networks, the cost function is taken to be the categorical cross-entropy and we use the Adadelta optimiser with adaptive learning rate (\citealt{adadelta}).  For our regression networks, our cost function is the mean squared error between the true and predicted values, and we use the Adam optimiser with a learning rate of 0.001 (\citealt{adam}).

The training data for each of our CNNs is described previously in Section \ref{sec:preptrain}.  Rather than passing an entire dataset into each network, we split up the data into many batches containing 128 samples each. Passing one of these batches through the full network allows us to evaluate the cost function which in turn allows us to minimise the cost with respect to the parameters of the network.  In this way, we iteratively adjust the trainable parameter values with each successive batch of training data.  A single training `epoch' is completed after all batches in the training data have been fed into the network.  We train each of our networks for 50 total epochs.

We construct our CNNs using the Keras API (\citealt{keras}) and the TensorFlow backend (\citealt{tensorflow}).  We train our networks with GPU acceleration using the same GeForce GTX 1080 Ti which was used to compute our N-body simulations.

\section{Results} \label{sec:res}

\subsection{Classifying S0 formation pathways} \label{sec:res1}

The CNNs which classify S0 {\bfd formation pathways} (i.e. models 1a, 1b, and 1c) yield a discrete class prediction for each input image.  This means that the accuracy of each trained network on a given dataset is straightforward to calculate by dividing the number of correct predictions by the total number of images in the dataset.

Our trained models 1a, 1b, and 1c each yield highly accurate predictions on the data with only marginal differences separating their performance.  Model 1a (which is trained on both morphological and kinematic images; see Table \ref{tab:nn}) provides the best results, with 99.8\% accuracy when predicting the S0 {\bfd formation pathway} for the training data, and 99.6\% accuracy for the test data.  Model 1b (which is trained on morphology alone) provides the lowest accuracies, but they are nevertheless extremely high at 99.0\% for the training data and 98.5\% for the test data.  Model 1c (which is trained on the kinematic data only) provides intermediate results, with 99.5\% accuracy for the training data and 99.1\% accuracy for the test data.

In summary, our trained CNNs provide remarkably accurate predictions for the {\bfd formation pathway} of our simulated S0s, no matter what dataset is used to train them.  Nevertheless, it seems kinematics convey marginally better information than morphology for the purposes of classification; but their combination provides the best results.

Incorrect classifications can be summarised in an error matrix, as shown in Table \ref{tab:errormat} for model 1a.  Rows indicate the true class for images in the given dataset and columns indicate the class which is predicted by the trained model.  Entries along the diagonal represent correct predictions, and off-diagonal entries are incorrect predictions.  The error matrix essentially tells us where the CNN struggles and where it is successful.  For instance, the entries equalling zero in Table \ref{tab:errormat} indicate that the CNN has no trouble distinguishing between the isolated and merger classes.  In other words, zero merger images are misclassified as the isolated class, and zero images of the isolated class are misclassified as mergers.

Using Table \ref{tab:errormat} we can compute other interesting quantities, such as the prediction accuracy for each S0 {\bfd formation pathway}.  For the training data, 99.9\% of the images in the merger class are predicted correctly, 99.7\% of the tidal class is predicted correctly, and 94.3\% of the isolated class is predicted correctly.  The corresponding accuracies on the test data are 99.9\%, 99.6\%, and 90.6\% for the merger, tidal, and isolated classes, respectively.  Because the accuracies on the training and test datasets are so similar, we can state with confidence that our CNNs do not suffer from overfitting.

The lowest accuracies occur for the isolated case.  The values of 94.3\% and 90.6\% for the training and test sets, respectively, are acceptable for the present work because they more than satisfy our goal of addressing our scientific questions.  For the future work described in Section \ref{sec:future}, however, we would seek to optimise the performance of our CNNs by improving these accuracies on the isolated class.

In Section \ref{sec:classfrac} we consider one factor which may explain why the predictions on the tidal and merger classes are superior: the isolated class is under-represented in the overall training data (comprising 1.75\% of the total) compared to the merger class (49.2\%) and tidal class (49.0\%).  Despite this lack of balance between the classes, the trained CNN performs remarkably well on the isolated class, with only $\approx 5 - 10\%$ misclassified as tidal, and zero misclassified as mergers.  The mild confusion between the tidal and isolated class may not be surprising because the morphology and kinematics produced in isolation may largely resemble those produced by weak tidal events.

\begin{table}
  \begin{center}
  \caption{Error matrices for the predicted S0 {\bfd formation pathways} (columns) compared to the {\bfd true pathways} (rows) after training our network on both the morphology and kinematics of the S0s, as described in the text.  Error matrices are given for the predictions on the training data (top) as well as the test data (bottom).  Entries along the diagonals show the number of correct classifications, while off-diagonal entries give the number and type of misclassifications.}

  \begin{tabular}{lccc}
  & \multicolumn{3}{c}{Predictions on training data} \\
  \cline{2-4}
  &&& \\
  & Isolated & Tidal & Merger \\

  True Isolated & 3537 & 214     & 0 \\
  True Tidal    & 17   & 104294 & 279 \\
  True Merger   & 0 	  & 3 	   & 104896 \\
  
  \hline

  & \multicolumn{3}{c}{Predictions on test data} \\
  \cline{2-4}
  &&& \\
  & Isolated & Tidal & Merger \\

  True Isolated & 950  & 99     & 0 \\
  True Tidal    & 5   & 25957  & 98 \\
  True Merger   & 0 	 & 15 	  & 26186 \\
  
  \hline

  \label{tab:errormat}
  \end{tabular}
  \end{center}
\end{table}

\subsection{Regression: predicting merger mass ratios} \label{sec:res2}

The CNNs which predict merger mass ratios (i.e. models 2a, 2b, 2c) do not permit a straightforward accuracy score as done previously for the classification CNNs.  This is because the predicted mass ratio for a given input image will be a continuous numerical value in the range $0.0 - 1.0$, and its discrepancy with respect to the true value must be addressed from a statistical perspective.

\begin{figure*}
   \begin{center}
   \includegraphics[width=0.95\textwidth]{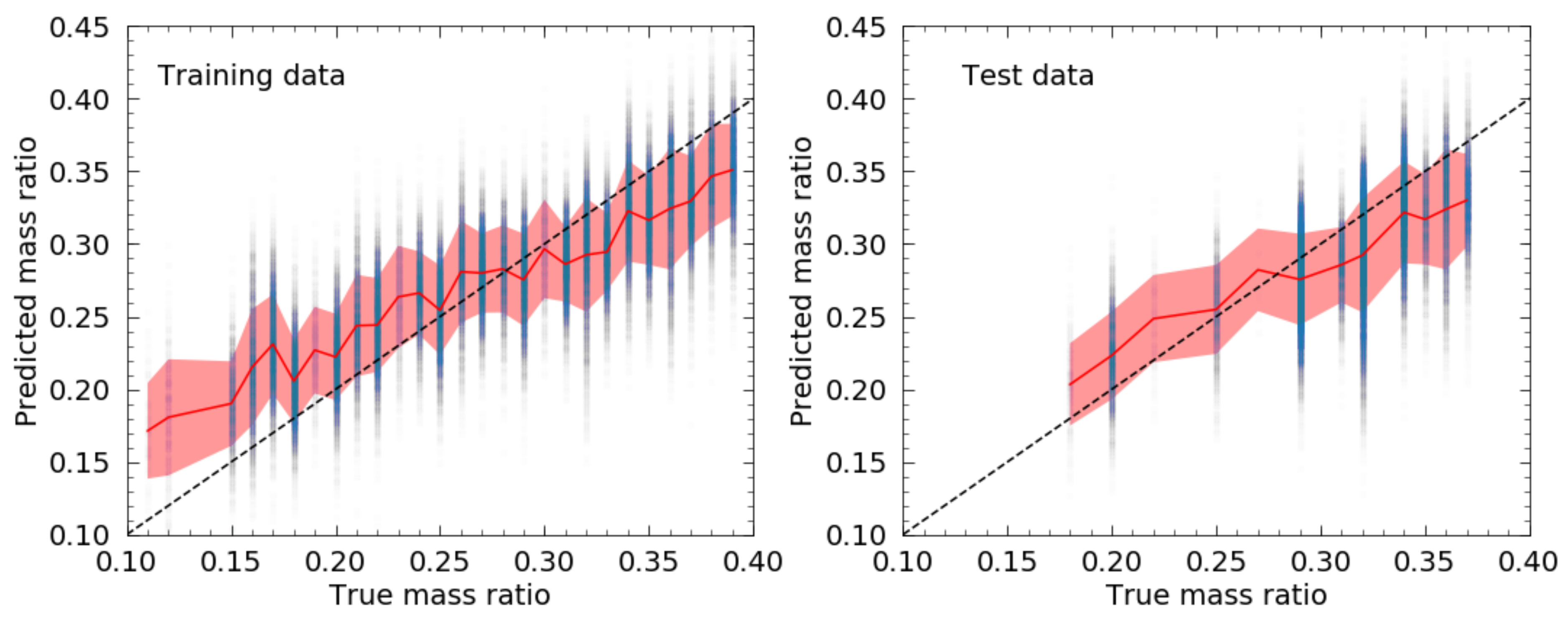}
   \caption{Predicted values for merger mass ratios versus their true values, where the predictions are generated by our trained CNN model 2a for images in the training dataset (left panel) and test dataset (right panel). The result for each image is represented by a faint blue circle in each panel, such that regions of overlap attain a dark blue color.  The red solid line gives the mean predicted value across the full range of true mass ratios.  The red shaded region indicates the $\pm 1 \sigma$ region around the mean, where $\sigma$ is computed as the standard deviation of all predicted values at each true value of the mass ratio.  The dashed black line is drawn as a visual aid to show where the predicted values perfectly match the true values.}
   \label{fig:merger_regression}
   \end{center}
\end{figure*}

Figure \ref{fig:merger_regression} displays the predicted mass ratios for model 2a on both the training dataset and test dataset.  The spread of predicted values at a given true value is nominally indicated by the scatter among the blue points, where each point represents the prediction for an individual image in the dataset.  The statistical spread is more accurately depicted by the red shaded region, which represents the $\pm 1 \sigma$ values centred on the mean predicted value for a given mass ratio.

The standard deviation in the predicted values is roughly constant $\approx 0.03$ across all true values for the mass ratios within both the training and test data.  This would seem to indicate that the predictions of the CNN are well-calibrated across the full dataset, and that the value of 0.03 represents a fundamental scatter within the features of the data independent of the mass ratio. 

In each panel, the red shaded region either encompasses or lies quite close to the dashed black line, which delineates an exact match between the predicted and true values.  This means that the mean predictions of model 2a are consistent with the true values to within $\approx 1 \sigma$ across the full range of mass ratios in the training and test datasets.  This gives us confidence that the trained network makes sensible predictions both in regions where there is an abundance of data (e.g. mass ratios of $0.25-0.35$) and in regions where there is relatively less data (e.g. for mass ratios of $0.15$ or less).

Discrepancies are nevertheless present and exhibit a clear trend, with the smallest mass ratios being over-predicted (by roughly 0.05) and the largest mass ratios being under-predicted (by roughly 0.03).  For the training data, the largest discrepancies occur at mass ratios of $0.11-0.12$, which may not be surprising because these mass ratios constitute the smallest fraction of the overall training dataset (e.g. see Figure \ref{fig:merger_hist}) {\bfd and thereby contribute the least to the training of the CNN.}

Also evident from Figure \ref{fig:merger_regression} is the fact that the predictions on the training data are quite similar to those on the test data (i.e. for the mass ratios within the test data).  Accordingly, we can conclude that our trained CNN does not suffer from overfitting.  As a side note, the full range of mass ratios is not present in the test data simply because of how we construct the training dataset, as described in Section \ref{sec:mehmrat}.  This difference between the two datasets is also indicated graphically in Figure \ref{fig:merger_hist}.

\begin{figure}
   \begin{center}
   \includegraphics[width=0.45\textwidth]{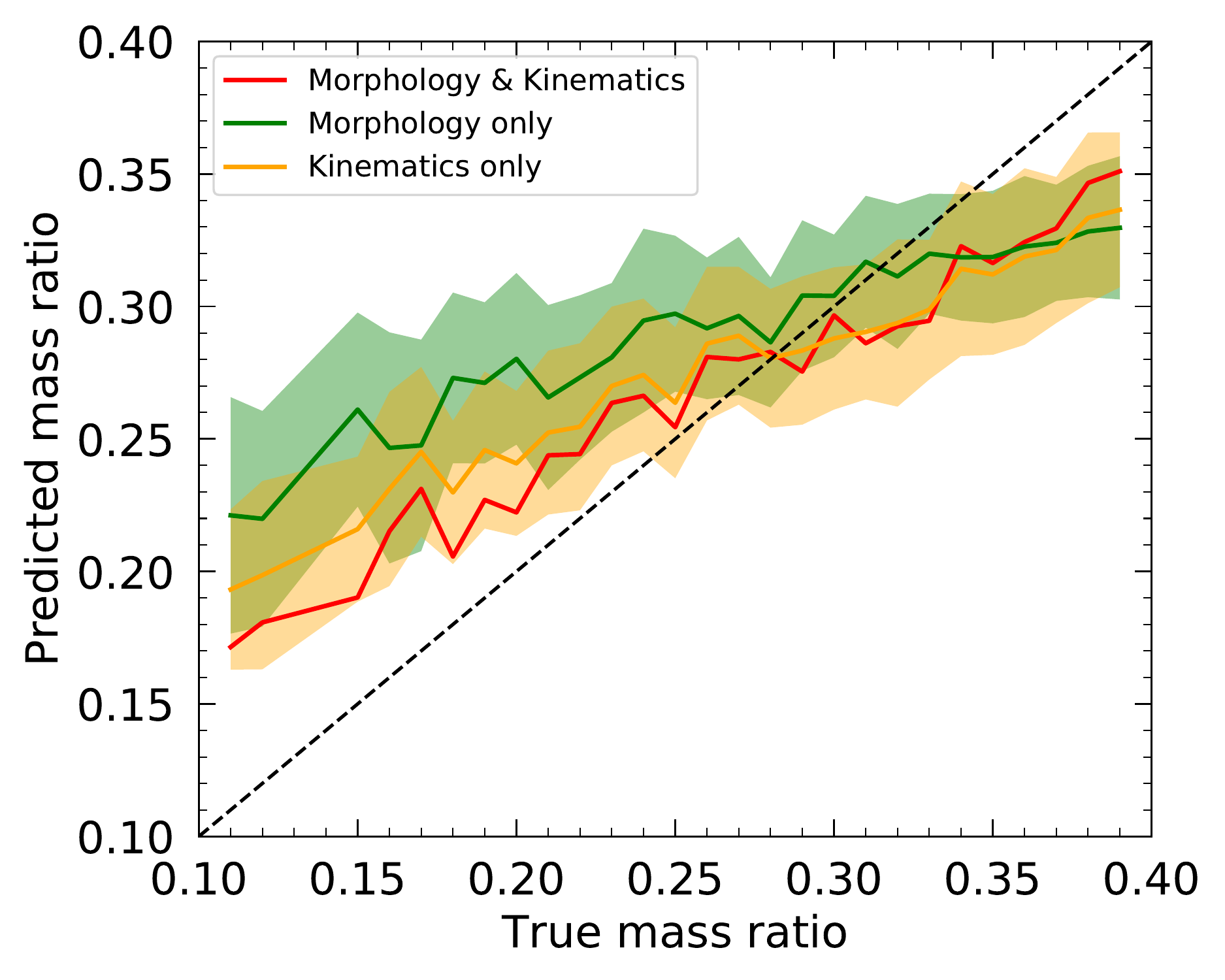}
   \caption{Predicted values for the merger mass ratios versus their true values, as predicted by three separate CNNs: model 2a (red), model 2b (green), and model 2c (yellow).  As summarised in Table \ref{tab:nn}, these CNNs are identical apart from the data used to train them: morphological and kinematic maps for model 2a, only morphology for model 2b, and only kinematic maps for model 2c. As in Figure \ref{fig:merger_regression}, the solid lines give the mean predicted values, and shaded regions give the $\pm 1 \sigma$ regions around the mean (the shaded region for model 2a is omitted for visual clarity).  Results here are shown for the training datasets only.  The solid red line in this plot is identical to the one in the left panel of Figure \ref{fig:merger_regression}.}
   \label{fig:merger_regression_compare_train}
   \end{center}
\end{figure}

Training the CNN on only morphology (model 2b) or only kinematic maps (model 2c) yields measurably different results.  Figure \ref{fig:merger_regression_compare_train} displays these differences, showing that the predictions for models 2b and 2c are inferior to those of model 2a across all mass ratios, with the mean discrepancies increasing in a manner which `flattens' the overall trend of predictions. The predictions when {\bfd trained} on morphology alone are particularly poor, with mean discrepancies exceeding 0.1 for small ratios, and with markedly increased dispersions about the mean.  This would suggest that S0 morphology conveys relatively little information on merger mass ratios.

In the range of mass ratios from $0.22$ to $0.37$, Figure \ref{fig:merger_regression_compare_train} indicates that the mean predictions of models 2a (red) and 2c (yellow) are broadly similar.  In other words, supplementing the kinematic maps with morphological data does not, on average, improve the performance of the CNN compared to training on kinematics alone for these mass ratios.  Beyond this range of mass ratios, however, the combination of morphological images with kinematic maps does yield an improvement in the predictions.  In other words, even though the morphology conveys comparatively little information on the mass ratio, this information nevertheless acts to complement the kinematic data and thereby improve the performance of the trained CNN.

\section{Discussion} \label{sec:discussion}

In this section, we discuss several issues including the effect of training our CNNs with different variations of the training datasets (e.g. changing the relative distributions of classes).  We also discuss how the predictions of our CNNs may depend mildly on the inclination of the S0 discs, and we sketch our plans for future work.

\subsection{Relative fraction of formation pathways} \label{sec:classfrac}

As described in Section \ref{sec:preptrain}, our training dataset for models 1a, 1b, and 1c is selected as a random 80\% subset of our overall data, with no consideration for preserving the relative fraction of each class.  For instance, the isolated class constitutes 1.80\% of the overall dataset, but the relative fraction of this class in the training data drops to 1.75\% due to random sampling.  In the test dataset, in contrast, the isolated class comprises 1.97\% of the total, as can be checked with the numbers given in Table \ref{tab:errormat}.

When we take a different random sample of the images as our training data, we can get marginally different results.  For instance, we repeated the training of our CNN model 1a with a different random training dataset for which the isolated class comprised 1.81\% of the total.  While this fractional increase may not seem very large, it is enough to produce a marked increase in the prediction accuracy for the isolated class.  This newly trained CNN predicts 98.7\% of the isolated class correctly in the training data, and 96.1\% correctly in the test dataset.  Recall the corresponding accuracies reported in our main results in Section \ref{sec:res1} were 94.3\% and 90.6\%, respectively.

This boost in performance of the CNN highlights the importance of optimising the relative fraction of classes within the data when training CNNs.  Ideally, each class should have equal representation in the data so that the training is not biased toward any particular class.  However, equal representation is not feasible in our case due to the nature of the isolated class.  That is, isolated formation histories will always be lacking in variety when compared to the diverse evolutionary histories possible via galactic interactions (e.g. {\bfd tidal and mergers}).  {\bfd Because our dataset aims to capture this evolutionary variety, the isolated class naturally constitutes a tiny fraction of the whole.

Nevertheless, in the future we can increase the number of isolated S0 models by varying structural and kinematic properties of the disc, such as the Toomre Q parameter.  By increasing the number of Spiral disc models, we will be able to have a larger variety of isolated S0 models and a larger number of corresponding images for training the CNNs.  As we have seen, such an increase in the fraction of training images  will likely improve the classification accuracies for the isolated formation pathway.}

\subsection{Distribution of mass ratios}

In Section \ref{sec:res2}, we report the results of using a roughly uniform distribution of merger mass ratios to train our models 2a, 2b, and 2c.  {\bfd We chose to construct the training data in this way in order to improve the overall accuracy for the CNNs across the full range of mass ratios.  When training the CNNs on a simple 80\% random sample of the dataset, the results are markedly inferior, particularly at low mass ratios.}  For instance, at a mass ratio of 0.11, this CNN overpredicts the true value by $0.1$ on average, and it overpredicts the mass ratios of 0.2 by $\sim 0.06$ on average.  This discrepancy is roughly a factor of two larger than the results shown in Figure \ref{fig:merger_regression} when training on the roughly uniform dataset.

Taking the training data as an 80\% random sample produces a dataset which is biased toward mass ratios of $0.29-0.32$, as can be seen in Figure \ref{fig:merger_hist}.  This range of values is notable because we find that the CNN trained with this biased training data exhibits its most accurate predictions in this same range, $0.29-0.31$.  This contrasts with our main results in Section \ref{sec:res2}, for which the most accurate predictions occur in the range $0.25-0.30$ as seen in Figure \ref{fig:merger_regression}.

Thus, when the training data is dominated by a particular subset of mass ratios, the performance of the CNN improves within that range of values but may suffer elsewhere, particularly at low mass ratios.  By taking a roughly uniform subset of the mass ratios, we are able to create a more balanced dataset.  (It nevertheless still suffers from under-representation of low mass ratios of $\approx 0.1$ as mentioned in Section \ref{sec:res2}.)  {\bfd Training with a balanced dataset improves the proportional representation of low mass ratios, and it likewise improves the predictions of the CNN at a wider range of input values.

Accordingly, we can further improve the accuracies of our CNNs by training with a truly uniform sample across the full range of mass ratios.  Even though we imposed an upper limit on the relative frequency at each mass ratio in our training data, our present dataset is still rather sparse below a mass ratio of 0.15 as seen in Figure \ref{fig:merger_hist}.  To assemble a truly uniform dataset, we would need to resimulate many more examples of S0 formation from mergers at low mass ratios with an expanded parameter space beyond our present investigation (e.g. Table \ref{tab:par}).  As this goes beyond the present scope, we leave this task to future study.}

\subsection{Morphology, kinematics, and their combination}

As described in Section \ref{sec:res1}, the CNNs which classify S0 {\bfd formation pathways} are more accurate when trained on kinematics (i.e. model 1c) than when trained on morphology (model 1b).  The advantages of the kinematic data are also apparent for the CNNs which predict merger mass ratios, because the performance of model 2c is markedly superior to model 2b (e.g. Figure \ref{fig:merger_regression_compare_train}).  This suggests that the physical imprints of S0 formation processes are best preserved in the kinematics rather than morphology.  This fact underscores the importance of {\bfd IFS surveys such as the SAMI galaxy survey (\citealt{croom2012}), SLUGGS (\citealt{brodie2014}), and MaNGA (\citealt{bundy2015})} to assemble the key data needed to illuminate the formation processes of observed S0s.

It is not surprising that our most accurate CNNs are trained on the combined dataset of morphology plus kinematics.  This is particularly evident in Figure \ref{fig:merger_regression_compare_train} for the CNNs which are trained to predict merger mass ratios, and it is true (albeit only marginally) for the CNNs which classify S0 {\bfd formation pathway} as reported in Section \ref{sec:res1}.  The reason for this is simply because combining multiple physical quantities into the data provides the largest set of independent inputs that can be used to tune the predictions of the network.

It is likely that bundling even more physical quantities into the training datasets will provide even better results.  For instance, one could imagine training the CNNs on image arrays with additional channels consisting of two-dimensional {\bfd velocity dispersion maps, metallicity maps, H-$\alpha$ distributions, neutral hydrogen kinematics, globular cluster kinematics, etc.}  Such data may help to improve the classification accuracies in areas where the CNN struggles, such as the mild confusion between the isolated class and tidal class described in Section \ref{sec:res1}.

Providing the CNNs with additional physical data during training may also reduce the statistical scatter in the predicted mass ratios.  It is likely that the scatter in Figure \ref{fig:merger_regression} can be traced to some extent to the diversity of morphologies and kinematic maps at a given value of the mass ratio, which in turn is influenced by the use of multiple initial conditions with varying bulge-to-disc ratios (Table \ref{tab:ic}).  By training the CNN on multiple additional physical quantities, the network has a better chance to extract the essential information to uniquely identify the mass ratio (or other physical measurements of interest) among the diversity of input images.

\subsection{Physical intuition for the CNN predictions}

\subsubsection{Feature Maps}

\begin{figure*}
   \begin{center}
   \includegraphics[width=0.95\textwidth]{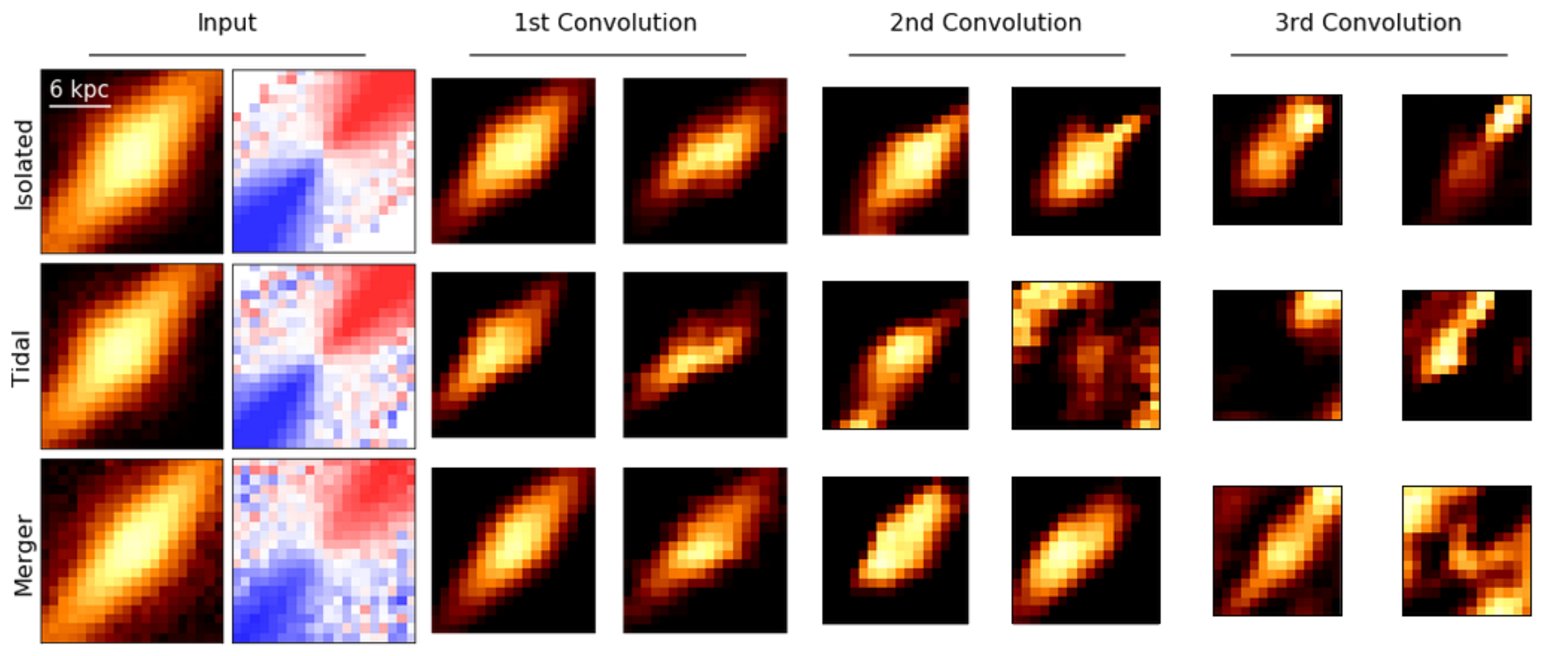}
   \caption{ {\bfd Three visually similar inputs shown alongside a selection of their feature maps for the classification of S0 formation pathways using our trianed Model 1a (see Table \ref{tab:nn}).  Each input comprises two images: stellar density (first column) and kinematic map (second column).  The physical scale and colormap are the same as in Figures \ref{fig:den} and \ref{fig:vel}.  We show one input for each formation pathway: isolated (top row), tidal (middle), and merger (bottom).  In each case, two feature maps are shown from each convolutional layer in the CNN, starting with the first convolutional layer (third and fourth columns), then the second convolutional layer (fifth and sixth columns), and then the third and final convolutional layer (seventh and eighth columns).  The feature maps were selected by choosing those with highest total activation (summed pixel value) after each layer.  For each of these inputs, the CNN assigns a probability exceeding 99.9\% to the correct formation pathway.} }
   \label{fig:featuremaps_classification}
   \end{center}
\end{figure*}

\begin{figure*}
   \begin{center}
   \includegraphics[width=0.95\textwidth]{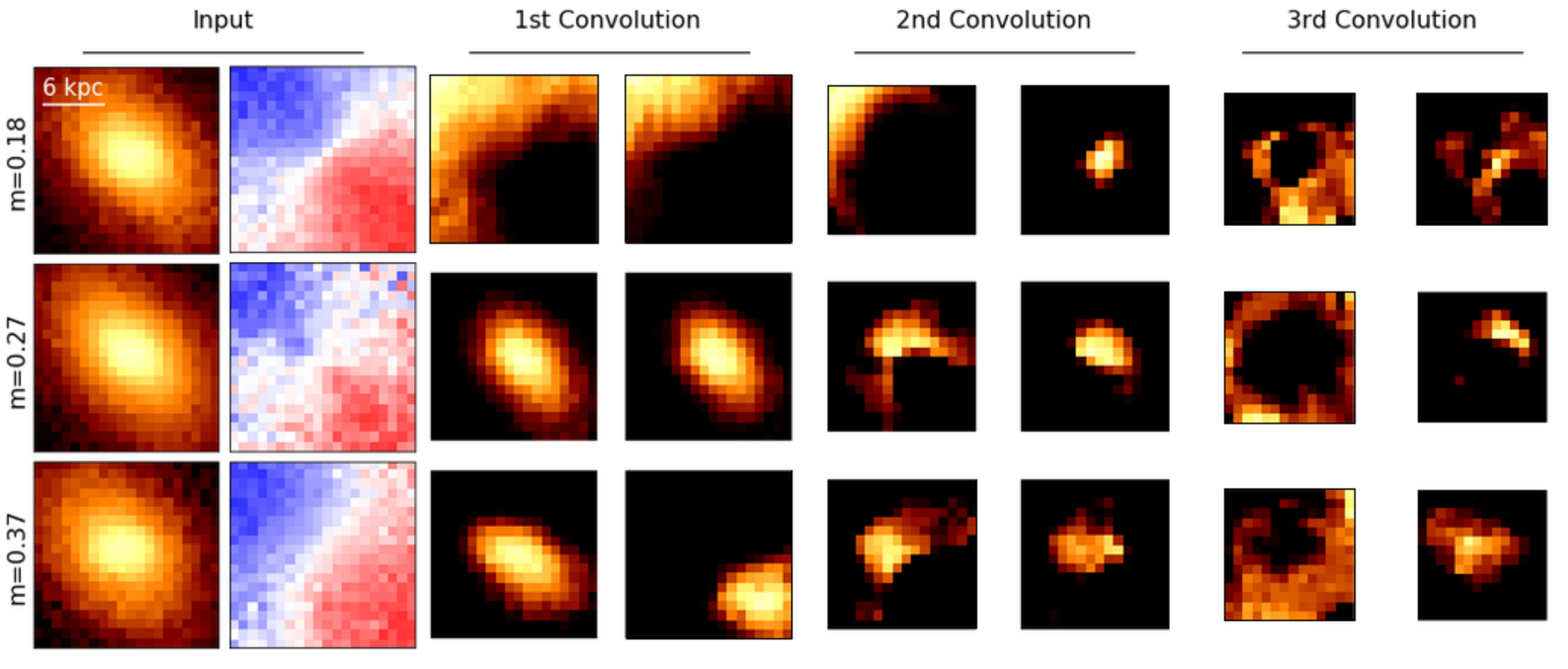}
   \caption{ {\bfd Three visually similar inputs shown alongside a selection of their feature maps for the prediction of merger mass ratios using our trianed Model 2a (see Table \ref{tab:nn}).  Each input comprises two images: stellar density (first column) and kinematic map (second column).  The physical scale and colormap are the same as in Figures \ref{fig:merger_den} and \ref{fig:merger_vel}.  We show one input from mass ratios of 0.18 (top row), 0.27 (middle), and 0.37 (bottom).  In each case, two feature maps are shown from each convolutional layer in the CNN, starting with the first convolutional layer (third and fourth columns), then the second convolutional layer (fifth and sixth columns), and then the third and final convolutional layer (seventh and eighth columns).  The feature maps were selected by choosing those with highest total activation (summed pixel value) after each layer.  For each of these inputs, the CNN correctly predicts the mass ratio to within an absolute error of 0.01.} }
   \label{fig:featuremaps_regression}
   \end{center}
\end{figure*}

{ \bfd We have demonstrated empirically in Section \ref{sec:res} that CNNs can accurately classify simulated S0s by their formation pathways and merger mass ratios.  However, we have not yet explored exactly how the CNN achieves this high performance, nor which physical features in the images are being utilised by the CNN.  To attempt to address these questions, we discuss in this section the so-called ``feature maps" of a selection of input images.

As an input image passes through the layers of the CNN, that image is transformed into many intermediate states called feature maps which the CNN utilises for its final prediction. These feature maps are the result of convolving one of the convolutional filters with the input to a layer.  When a single input image is passed into our CNN (see the architecture schematic in Figure \ref{fig:nn}), the first convolutional layer produces 32 feature maps (each of size $18 \times 18$ pixels), the second convolutional layer produces 64 feature maps (of size $16 \times 16$), and the final convolution produces 128 feature maps (of size $14 \times 14$).

In Figure \ref{fig:featuremaps_classification} we show feature maps pertaining to three visually similar input images from each formation pathway.  In each case, the CNN correctly predicts the formation pathway with probability exceeding 99.9\%.  Owing to limitations of space, it is not feasible to inspect all feature maps for a given input image, so instead we restrict our attention to the two feature maps with highest total activation (i.e. highest summed pixel value) from each layer.

The feature maps from the first convolutional layer (columns three and four in Figure \ref{fig:featuremaps_classification}) are all quite similar, which suggests that the CNN is not able to distinguish the three formation pathways after only the first convolution.  The feature maps begin to diverge visually after the second convolutional layer (columns five and six), and the variety is even larger after the third convolution (final two columns).  This highlights the fact that a CNN requires numerous convolutional layers stacked together to be able to extract meaningful feature maps from the input.

Focusing on the final two columns of Figure \ref{fig:featuremaps_classification}, we can infer that most of the feature maps have high activation (high pixel value) in the midplane of the disc where the rotational velocities likely peak.  This suggests that the CNN is using rotational amplitudes in some way to distinguish the pathways.  The emphasis of the central region in the feature maps varies between the three pathways, which may signify the varying importance of the bulge across the three progenitor models we consider.  The feature map in the bottom right corner of Figure \ref{fig:featuremaps_classification} is quite interesting, as it exhibits activations across broad regions of the inner disc and above the disc plane.  This likely means the CNN is picking up on the presence of a greater amount of random motions for mergers in comparison to the other pathways.

We specifically selected input images which appear similar in morphology and kinematics in Figure \ref{fig:featuremaps_classification} in order to highlight the following point: the human eye is almost certainly unable to distinguish the true formation pathways due to the similarity of these images, but the CNN predicts the correct pathway with a probability exceeding 99.9\%.
The same is true for Figure \ref{fig:featuremaps_regression}, which shows three similar inputs to our second CNN for predicting merger mass ratios.  The CNN again exhibits great performance despite the similarity in the images, predicting the correct mass ratio in each case to within an absolute error of 0.01.

Whereas the feature maps of Figure \ref{fig:featuremaps_classification} primarily emphasize the elongated and rapidly rotating disc, the feature maps of Figure \ref{fig:featuremaps_regression} emphasize different regions.  
For instance, the feature maps of the final convolutional layer (final two columns in Figure \ref{fig:featuremaps_regression}) strikingly emphasize a broad region of the disc while excluding the inner region where the bulge dominates, creating a visual impression of a hole.  This suggests that the key information needed to infer merger mass ratios may be preserved in the disc rather than the bulge, and may be connected to how much the galactic disc has been disturbed following the merger.

\subsubsection{The role of randomness} \label{sec:random}

Inspecting feature maps can provide hints at how the CNN operates, but in general there is no straightforward way to extract a simple physical interpretation.  This is particularly important to remember when inspecting feature maps of only a few input images as in Figures \ref{fig:featuremaps_classification} and \ref{fig:featuremaps_regression}.  It is tempting to isolate differences in the properties of these few images and draw conclusions for physical intuition, but the network is not producing feature maps based on the differences between just these few images; it is trained to distinguish the differences between the full set of training data, comprising on the order of one hundred thousand input images.  Inspecting the full set of feature maps of all input data is simply not feasible.

Instead, we may speculate on the broad properties that distinguish the images of one formation pathway from another.  In particular, the amount of pixel-by-pixel randomness in both density images and kinematic maps seems to be correlated with formation pathway.  Here we use the term `randomness' to qualitatively denote areas of an image which exhibit large fluctuations in the amplitude of velocity (for the kinematic maps) or density (for the morphological maps), such that the image does not appear smooth.  For the images of the present study, the presence of such randomness is an intrinsic feature of the simulations themselves rather than a product of how the images are constructed.

When inspecting various images (e.g. Figures \ref{fig:den}, \ref{fig:vel}, and \ref{fig:featuremaps_classification}), one may judge that the least amount of randomness is present in the isolated images and the greatest amount of randomness appears in the merger images.  This seems to hold true regardless of the size of the initial bulge in the Spiral model.  The presence of randomness is somewhat more clear within the kinematic maps in comparison to the density maps, for several possible reasons.  As a vector quantity, the velocities of different particles can sum in opposite directions whereas density maps are summed from scalar masses.  This may naturally allow the amplitude of random motions to be stronger than the amplitude of random density variations.  Also, the logarithmic scaling of the density images suppresses the scale of any randomness which is present.

There are physical reasons why we would expect S0s to exhibit different degrees of randomness for each of our three pathways. An essential ingredient to the formation of S0s is the disappearance of spiral features due to disc heating, but the mechanism for dynamical heating can vary in terms of strength and direction for different pathways.  It is least disruptive for the isolated pathway, wherein clumps migrate through the disc (but never out of the disc) and produce an overall smooth velocity field.  The heating mechanism for mergers is the most violent (strong mergers can destroy the disc altogether) and can occur in random directions based on the relative orientations of the two galaxies.  The tidal pathway stands in between, as the tidal forces are strong enough to warp and disrupt a disc though not destroy it, and the direction may be randomly oriented with respect to the initial disc.  Accordingly it makes intuitive sense that the amount of pixel-to-pixel randomness in the images is likely correlated with the disc heating mechanism, with progressively larger degrees of randomness for the isolated, tidal, and merger pathways, respectively.

If the above intuition holds true, then we may further speculate that the CNN is able to capture these subtle differences in image randomness and thereby produce highly accurate predictions.  This may also explain why the CNN struggles in some cases. For instance, the large errors in the predictions at low merger mass ratios may potentially be explained by the relative lack of randomness for low mass ratio mergers in comparison to higher mass ratio mergers.

Though the discussion is only qualitative at this point, we will seek to quantify the amount of such randomness present in our training data in the future.  More importantly, we plan to correlate the predictions of the CNNs with the amount of randomness or other potentially important quantities.  In this way we may be able to address the key question of how the CNN generates its predictions on our dataset and thereby understand how to best improve the results. 

}

\subsection{Accuracies as a function of inclination}

\begin{figure*}
   \begin{center}
   \includegraphics[width=0.95\textwidth]{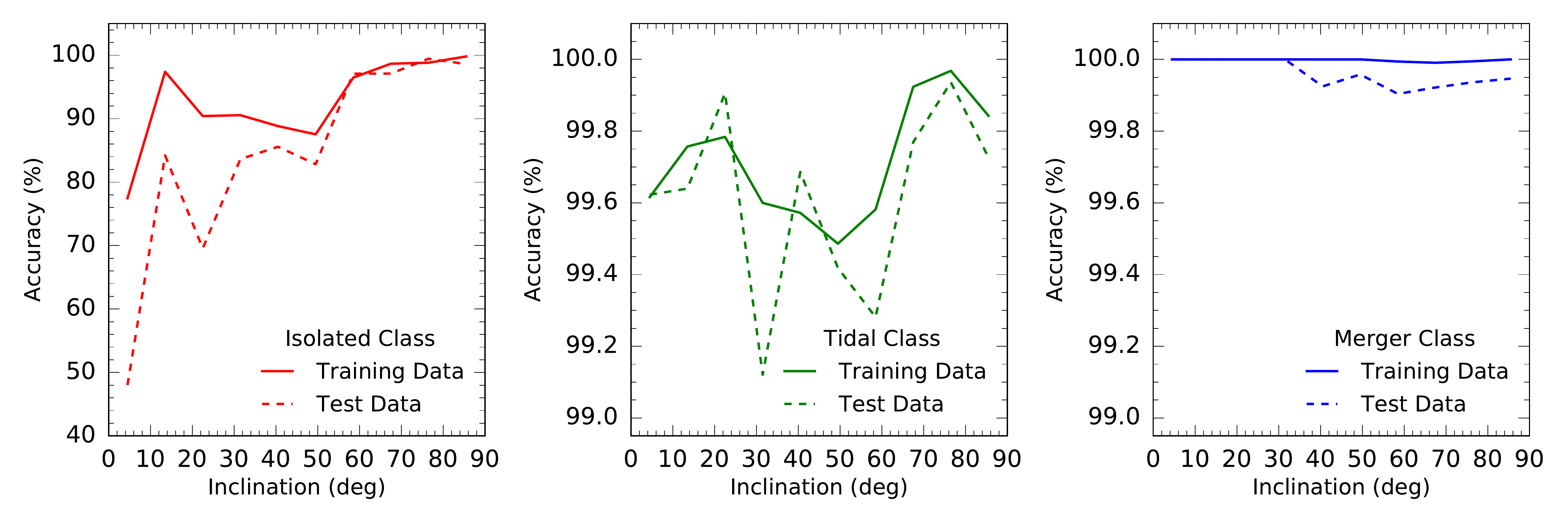}
   \caption{Prediction accuracies of the trained CNN model 1a as a function of galactic inclination for a given set of input images.  Each panel provides the accuracies for a given S0 {\bfd formation pathway}: isolated (left), tidal (middle), and merger (right).  Solid lines indicate the predictions on the training dataset, and dashed lines pertain to the test set.  The accuracies for the tidal and merger class exceed 99\% for all inclinations.  In contrast, the accuracies for the isolated class reach as low as approximately 50\% and 80\% for face-on images in the test and training data, respectively (note the different vertical range in the left panel).}
   \label{fig:inc_classification}
   \end{center}
\end{figure*}

\begin{figure}
   \begin{center}
   \includegraphics[width=0.45\textwidth]{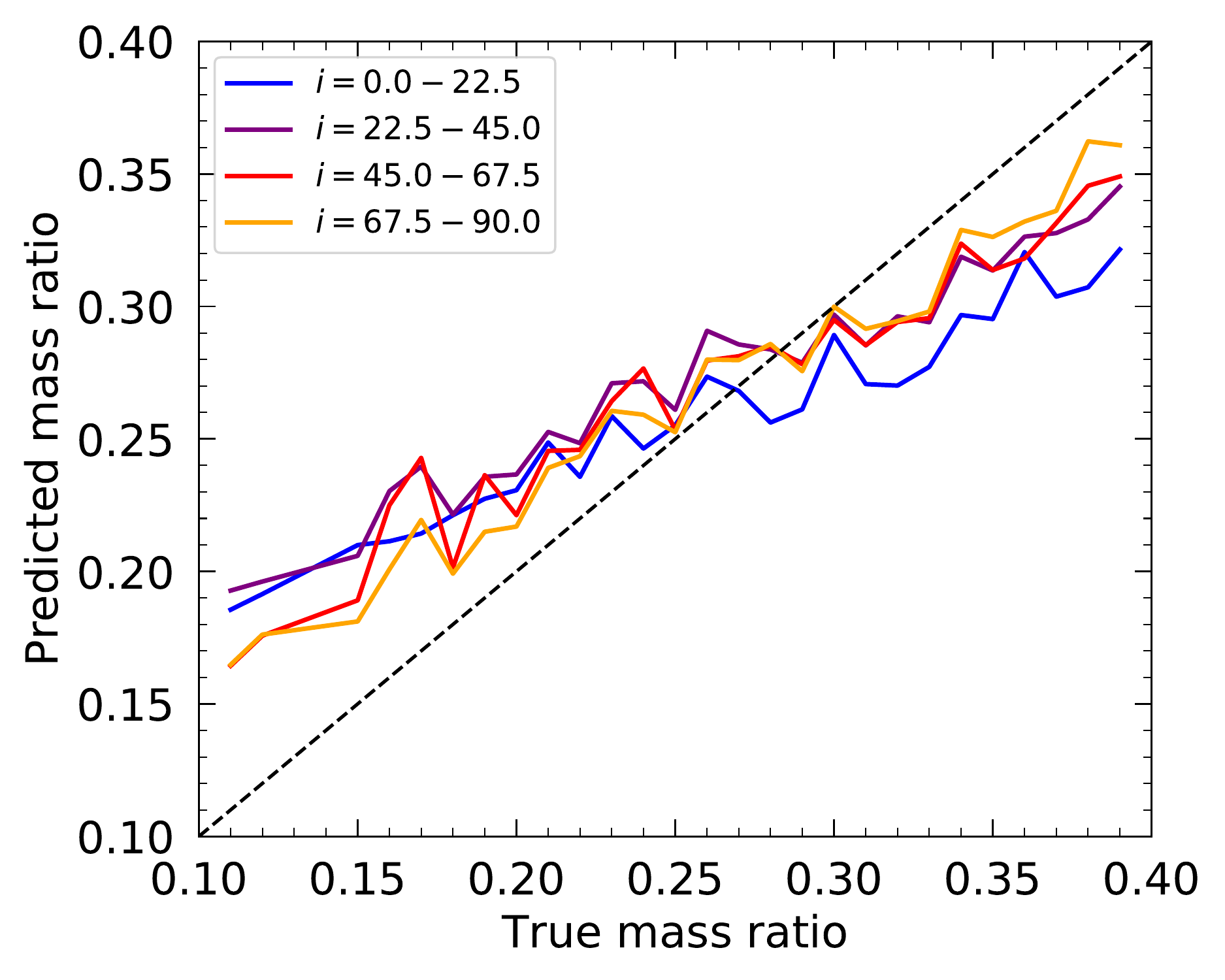}
   \caption{Mean predicted values for the merger mass ratio as given by our trained CNN model 2a on four subsets of the training data: S0s with inclinations of $0\deg-22.5\deg$ (blue), $22.5\deg-45\deg$ (purple), $45\deg-67.5\deg$ (red), and $67.5\deg-90\deg$ (yellow).}
   \label{fig:inc_regression}
   \end{center}
\end{figure}

{\bfr Here we consider whether the accuracy of our CNNs is affected by the inclination of the S0 discs (e.g. whether face-on or edge-on orientations are better classified). We emphasise that the projected images in our synthetic dataset vary in terms of disc inclination as well as the azimuthal angle of the disc. However, our simulated S0s are generally axisymmetric, which implies the variation of azimuthal angle is not as important as the variation of inclination.}

Figure \ref{fig:inc_classification} shows the prediction accuracy of our trained CNN model as a function of inclination for each S0 {\bfd formation pathway}.  The left panel indicates a noticeable trend for the isolated class: the accuracies generally increase with disc inclination, for both the training and test datasets.  The accuracies drop to $\approx 50\%$ for face-on S0s in the test data, and $\approx 80\%$ for face-on images of the training data.

Such a significant drop in accuracy may be due in part to the small total number of images in the isolated class (e.g. see Section \ref{sec:classfrac}).  Nevertheless it likely implies that the edge-on images of the isolated class contain more distinguishing features in comparison to the face-on images.  This may be surprising, because an edge-on image contains fewer non-trivial pixels than face-on, simply because the sky background covers more of the image.  In other words, despite the fact that edge-on galaxies span fewer pixels, these orientations are easier for the CNN to classify correctly (for the isolated class).

There may be several physical reasons for this.  For instance, face-on kinematic maps convey relatively little information because the orbits within the disc are aligned orthogonally to the image projection.  Thus, the salient kinematic features of each S0 {\bfd formation pathway} may largely disappear in the face-on projection, making such images harder to classify.  In addition, each of the galaxies is roughly axisymmetric, which suggests that the features of the disc may be largely redundant as a function of azimuthal angle.  Lastly, the sharp sky background of the edge-on isolated images may prove useful to the CNN, particularly when the tidal and merger scenarios produce discs with increased vertical scale-heights and dispersions, which may be easily detected in edge-on but not face-on projections.

Trends with inclination are less noticeable for the tidal (middle panel) and merger classes (right) in Figure \ref{fig:inc_classification}.  This is primarily because the accuracies are extremely high for all inclinations, exceeding 99.1\% and 99.9\%, respectively.  The tidal class may exhibit a weak upward trend with inclination for the training data, but the same cannot be said for the test data.  The merger class, meanwhile, shows its lowest accuracies for higher inclinations, both for the training and test data.  However, this trend does not signify a systemic issue for the class because vanishingly few images are misclassified.

Figure \ref{fig:inc_regression} shows the mean prediction for merger mass ratios for a set of different inclinations as given by our trained model 2a.  For this CNN, the predictions are most accurate at high inclinations (yellow line), and the predictions seem to be least accurate at low inclinations (blue line), particularly at high mass ratios (e.g. $> 0.25$).  At low mass ratios ($<0.25$), the mean predictions are roughly equivalent for different inclinations, but the highest inclinations nevertheless still perform the best.  The physical reason underlying these trends is likely related to the information contained in the vertical thickness and dispersion of the discs, which is amplified in particular for high mass ratio mergers.

It is interesting that the right panel of Figure \ref{fig:inc_classification} indicates an opposite trend with inclination compared to that of Figure \ref{fig:inc_regression}.  In other words, the classification of merger images by {\bfd formation pathway} has highest accuracy at low inclinations; but the prediction of merger mass ratios has highest accuracy at high inclinations.  Such a contrast highlights the fact that the quality of information contained in a galactic image can change depending on the physical quantity being measured.

\subsection{Future Work} \label{sec:future}

Though we have focused exclusively on simulated data in this paper, our ambitions for future work are to extract fossil information from observational data.  In particular, our strategy is to train CNNs using the known formation processes of simulated S0s and to subsequently feed observed images into these pre-trained networks for classification.  In other words, we will be able to predict the fundamental physical quantities which underpin each observed system.  The current work serves to validate this approach by demonstrating its viability on a simulated dataset.

The success of our future work depends on a variety of factors.  For instance, the simulated images and kinematic maps must be produced with a greater level of sophistication than in the present work, so that they are fully representative of the features contained in the observational data.  {\bfd This includes, for instance, adding realistic image noise to mimic observational conditions as well as choosing appropriate image resolutions and scales.} In addition, the variety of {\bfd formation pathways} in the simulations must equal or exceed the true range of formation histories of the observed systems.  Because the true formation histories are unknown, we can attempt to satisfy this requirement by creating a substantially larger set of simulations with an expanded parameter space.  It may also be important to incorporate additional physics not considered in this work, such as hydrodynamical interactions with gaseous halos, cosmological gas accretion, galactic feedback, etc.

\section{Summary and Conclusion} \label{sec:summary}

In this paper we have demonstrated that CNNs can be trained to {\bfd classify simulated S0 galaxies according to their formation pathways} and merger mass ratios.  {\bfr We first assembled a large set of N-body simulations which model the formation of S0s from Spiral progenitors under idealised but reasonable conditions for the merger, tidal, and isolated pathways.  We then created a dataset of stellar density images and two-dimensional kinematic maps of our simulated S0s, and we trained our CNNs on these images.}  Our trained CNNs successfully {\bfd confirm each S0 formation pathway} for the vast majority of images across various training and test datasets.  We also find success when predicting the merger mass ratios, as our trained CNNs produce mean predictions which differ from the true values to within roughly $1 \sigma$ across the full range of our data.

{\bfd In the future, we will apply our trained networks to observed images and kinematic maps of S0s, and thereby classify observed systems based on fundamental physical processes.  In principle, the applicability of our method is not limited to the classification of S0s.  For instance, the methodology could be applied to other types of galaxies with multiple formation pathways, such as dwarf ellipticals which may form by mergers or by tidal stirring (\citealt{mayer2001,yozin2012}).}

Overall, we find that the CNNs trained on kinematic maps are more accurate than those trained on only images of the stellar density; but the most accurate CNNs are trained on the combination of kinematics plus morphology.  This is particularly true for predictions of merger mass ratios.  This suggests that the imprints of fundamental formation processes are {\bfd found primarily} in the kinematics of a galaxy, which in turn highlights the importance of observational surveys to assemble the kinematic data needed to illuminate such processes in the real universe. {\bfd Key questions remain to be answered, such as clarifying exactly how the CNNs achieve their accurate predictions and why the kinematic data are superior to the density maps.}

The utility of training CNNs to achieve our scientific goals is particularly relevant in light of the remarkable similarities among the morphologies and kinematics of the input data.
Due to these similarities, it is likely quite difficult if not impossible for astronomers to visually predict formation processes with accuracies that would compete with the CNNs.  This fact underscores the role that CNNs may provide in the future to augment the abilities of astronomers.

In summary, our current work is an important step toward classifying galaxies by the fundamental physical properties which drive their evolution.  We conclude that training CNNs on simulated datasets of S0 formation may provide key insights in the future toward dividing S0s into an assortment of physically-motivated subcategories.

\section*{Acknowledgements}

This research was supported by the Australian government through the Australian Research Council's Discovery Projects funding scheme (DP170102344).

\bibliographystyle{mnras}

\begin{thebibliography}{}

\bibitem[Abadi et al.(2016)]{tensorflow} Abadi, M., Agarwal, A., Barham, P., et al.\ 2016, arXiv:1603.04467 

\bibitem[Ackermann et al.(2018)]{ackerman2018} Ackermann, S., Schawinski, K., Zhang, C., Weigel, A.~K., \& Turp, M.~D.\ 2018, \mnras, 479, 415 

\bibitem[Barway et al.(2013)]{barway2013} Barway, S., Wadadekar, Y., Vaghmare, K., \& Kembhavi, A.~K.\ 2013, \mnras, 432, 430

\bibitem[Bekki(1998)]{bekki98} Bekki, K.\ 1998, \apjl, 502, L133 

\bibitem[Bekki \& Couch(2011)]{bekkicouch2011} Bekki, K., \& Couch, W.~J.\ 2011, \mnras, 415, 1783 

\bibitem[Bekki(2013)]{bekki2013} Bekki, K.\ 2013, \mnras, 432, 2298 

\bibitem[Bekki(2014)]{bekki2014} Bekki, K.\ 2014, \mnras, 444, 1615 

\bibitem[Bekki(2015)]{bekki2015} Bekki, K.\ 2015, \mnras, 449, 1625 

\bibitem[Bekki(2019)]{bekki2019} Bekki, K.\ 2019, \mnras, 485, 1924 

\bibitem[Bekki et al.(2019)]{aiverse} Bekki, K., Diaz, J.~D., \& Stanley, N.\ 2019, Astronomy and Computing, in press.

\bibitem[\protect\citeauthoryear{{Binney} \& {Tremaine}}{{Binney} \& {Tremaine}}{2008}]{BT}
{Binney} J.,  {Tremaine} S.,  2008, {Galactic Dynamics: Second Edition}.
Princeton University Press

\bibitem[Brodie et al.(2014)]{brodie2014} Brodie, J.~P., Romanowsky, A.~J., Strader, J., et al.\ 2014, \apj, 796, 52 

\bibitem[Bryant et al.(2015)]{bryant2015} Bryant, J.~J., Owers, M.~S., Robotham, A.~S.~G., et al.\ 2015, \mnras, 447, 2857 

\bibitem[Bruzual \& Charlot(2003)]{bruzual2003} Bruzual, G., \& Charlot, S.\ 2003, \mnras, 344, 1000 

\bibitem[Bundy et al.(2015)]{bundy2015} Bundy, K., Bershady, M.~A., Law, D.~R., et al.\ 2015, \apj, 798, 7 

\bibitem[Buta et al.(2007)]{buta2007} Buta, R.~J., Corwin, H.~G., \& Odewahn, S.~C.\ 2007, The de Vaucouleurs Atlas of Galaxies, edited by Ronald J.~Buta, Harold G.~Corwin and Stephen C.~Odewahn.~ISBN-13 978-521-82048-6 (HB).~Published by Cambridge University Press, Cambridge, UK, 2007.  

\bibitem[Byrd \& Valtonen(1990)]{byrd90} Byrd, G., \& Valtonen, M.\ 1990, \apj, 350, 89 

\bibitem[Cappellari et al.(2011)]{cap2011} Cappellari, M., Emsellem, E., Krajnovi{\'c}, D., et al.\ 2011, \mnras, 413, 813 

\bibitem[Charnock \& Moss(2017)]{charnock2017} Charnock, T., \& Moss, A.\ 2017, \apjl, 837, L28 

\bibitem[Chollet et al. (2015)]{keras} Chollet, Fran\c{c}ois, et al. 2015, https://keras.io

\bibitem[Croom et al.(2012)]{croom2012} Croom, S.~M., Lawrence, J.~S., Bland-Hawthorn, J., et al.\ 2012, \mnras, 421, 872 

\bibitem[Dai \& Tong(2018)]{dai2018} Dai, J.-M., \& Tong, J.\ 2018, arXiv:1807.10406 

\bibitem[de Vaucouleurs(1959)]{devauc59} de Vaucouleurs, G.\ 1959, Handbuch der Physik, 53, 275

\bibitem[Diaz et al.(2018)]{diaz2018} Diaz, J., Bekki, K., Forbes, D.~A., et al.\ 2018, \mnras, 477, 2030 

\bibitem[Dieleman et al.(2015)]{dieleman2015} Dieleman, S., Willett, K.~W., \& Dambre, J.\ 2015, \mnras, 450, 1441 

\bibitem[Dom{\'{\i}}nguez S{\'a}nchez et al.(2018)]{ds2018} Dom{\'{\i}}nguez S{\'a}nchez, H., Huertas-Company, M., Bernardi, M., Tuccillo, D., \& Fischer, J.~L.\ 2018, \mnras, 476, 3661 

\bibitem[Forbes et al.(2016)]{forbes2016} Forbes, D.~A., Romanowsky, A.~J., Pastorello, N., et al.\ 2016, \mnras, 457, 1242 

\bibitem[Fraser-McKelvie et al.(2018)]{fm2018} Fraser-McKelvie, A., Arag{\'o}n-Salamanca, A., Merrifield, M., et al.\ 2018, arXiv:1809.04336 

\bibitem[Graham \& Worley(2008)]{gw2008} Graham, A.~W., \& Worley, C.~C.\ 2008, \mnras, 388, 1708 

\bibitem[Hoyle(2016)]{hoyle2016} Hoyle, B.\ 2016, Astronomy and Computing, 16, 34 

\bibitem[Hubble(1936)]{hubble36} Hubble, E.~P.\ 1936, Realm of the Nebulae, by E.P.~Hubble.~ New Haven: Yale University Press, 1936.~ ISBN 9780300025002  

\bibitem[Kennicutt(1998)]{kennicutt98} Kennicutt, R.~C., Jr.\ 1998, \araa, 36, 189 

\bibitem[Kingma \& Ba(2014)]{adam} Kingma, D.~P., \& Ba, J.\ 2014, arXiv:1412.6980 

\bibitem[Kormendy(1977)]{kormendy77} Kormendy, J.\ 1977, \apj, 218, 333 

\bibitem[Kormendy \& Bender(2012)]{kormendy2012} Kormendy, J., \& Bender, R.\ 2012, \apjs, 198, 2 

\bibitem[Kroupa(2001)]{kroupa2001} Kroupa, P.\ 2001, \mnras, 322, 231 

\bibitem[Laurikainen et al.(2010)]{lauri2010} Laurikainen, E., Salo, H., Buta, R., Knapen, J.~H., \& Comer{\'o}n, S.\ 2010, \mnras, 405, 1089

\bibitem[Ma et al.(2014)]{ma2014} Ma, C.-P., Greene, J.~E., McConnell, N., et al.\ 2014, \apj, 795, 158 

\bibitem[Mayer et al.(2001)]{mayer2001} Mayer, L., Governato, F., Colpi, M., et al.\ 2001, \apj, 559, 754 

\bibitem[Navarro et al.(1996)]{navarro96} Navarro, J.~F., Frenk, C.~S., \& White, S.~D.~M.\ 1996, \apj, 462, 563 

\bibitem[Neto et al.(2007)]{neto2007} Neto, A.~F., Gao, L., Bett, P., et al.\ 2007, \mnras, 381, 1450 

\bibitem[Noguchi(1998)]{noguchi98} Noguchi, M.\ 1998, \nat, 392, 253 

\bibitem[Petrillo et al.(2018)]{petrillo2018} Petrillo, C.~E., Tortora, C., Chatterjee, S., et al.\ 2018, arXiv:1807.04764 

\bibitem[Prieto et al.(2013)]{prieto2013} Prieto, M., Eliche-Moral, M.~C., Balcells, M., et al.\ 2013, \mnras, 428, 999 

\bibitem[Quilis et al.(2000)]{quilis2000} Quilis, V., Moore, B., \& Bower, R.\ 2000, Science, 288, 1617  

\bibitem[Saha \& Cortesi(2018)]{saha2018} Saha, K., \& Cortesi, A.\ 2018, \apjl, 862, L12 

\bibitem[S{\'a}nchez et al.(2012)]{sanchez2012} S{\'a}nchez, S.~F., Kennicutt, R.~C., Gil de Paz, A., et al.\ 2012, \aap, 538, A8 

\bibitem[Sandage(1961)]{sandage61} Sandage, A.\ 1961, The Hubble Atlas of Galaxies, Washington: Carnegie Institution, Publ. No. 618

\bibitem[Shah et al.(2019)]{shah2019} Shah, M., Bekki, K., Vinsen, K., \& Foster, S.\ 2019, \mnras, 482, 4188 

\bibitem[Stark et al.(2018)]{stark2018} Stark, D., Launet, B., Schawinski, K., et al.\ 2018, \mnras, 477, 2513 

\bibitem[Tsujimoto et al.(1995)]{tsujimoto95} Tsujimoto, T., Nomoto, K., Yoshii, Y., et al.\ 1995, \mnras, 277, 945 

\bibitem[Tuccillo et al.(2018)]{tuccillo2018} Tuccillo, D., Huertas-Company, M., Decenci{\`e}re, E., et al.\ 2018, \mnras, 475, 894 

\bibitem[van den Hoek \& Groenewegen(1997)]{vdh97} van den Hoek, L.~B., \& Groenewegen, M.~A.~T.\ 1997, \aaps, 123, 305 

\bibitem[Yozin \& Bekki(2012)]{yozin2012} Yozin, C., \& Bekki, K.\ 2012, \apjl, 756, L18 

\bibitem[Zeiler(2012)]{adadelta} Zeiler, M.~D.\ 2012, arXiv:1212.5701 

\end{thebibliography}

\appendix

\section{Details of adopted simulation code} \label{sec:code}

{ \bfr

Here we discuss various implementation details of our simulation code as it relates to our present study of S0s.  For full details on the code, see \citet{bekki2013} (hereafter B13) and \citet{bekki2014}.

A range of physical processes are modelled self-consistently in our simulations, including: gas dynamics (smoothed-particle hydrodynamics), star formation, ${\rm H_2}$ formation on dust grains, formation of dust grains in the stellar winds of supernovae and asymptotic giant branch stars, time evolution of interstellar radiation field, growth and destruction processes of dust in the interstellar medium, and ${\rm H_2}$ photo-dissociation due to far ultra-violet light. The code does not include feedback from active galactic nuclei on the interstellar medium nor the growth of supermassive black holes. Such feedback effects could be important for several pathways of S0 formation (particularly in relation to central star formation), so we will investigate such effects in the future.

Applied to the formation of S0s, our numerical code allows us to derive the structural and  kinematical properties, dust abundances, and spatial distributions of atomic and molecular hydrogen.  For the present work, however, we focus exclusively on the stellar properties of our S0 models in order to draw parallels with IFS surveys.

We adopt the `${\rm H_2}$-dependent' star formation recipe of B13 in which the star formation rate (SFR) is determined by the local molecular fraction ($f_{\rm H_2}$) of each gas particle.  A gas particle {\it can be} converted into a new star if the following three conditions are met: (i) the local dynamical time scale is shorter than the sound crossing time scale (mimicking the Jeans instability), (ii) the local velocity field is identified as being consistent with gravitational collapse (i.e., $\nabla \cdot {\bf v}<0$), and (iii) the local density exceeds the threshold density of 1 cm$^{-3}$ for star formation.  We also adopt
the  Kennicutt-Schmidt law, which is described as $ {\rm SFR} \propto \rho_{\rm g}^{\alpha_{\rm sf}}$;  (\citealt{kennicutt98}),
where $\alpha_{\rm sf}$ is the power-law slope. A reasonable value of $\alpha_{\rm sf}=1.5$ is adopted for all models.

Each supernova (SN) is assumed to eject a total feedback energy ($E_{\rm sn}$) of $10^{51}$ erg.  Of this total, 90\% and 10\% of $E_{\rm sn}$ are deposited as an increase of thermal energy (`thermal feedback') and random motion (`kinetic feedback'), respectively. The thermal energy is used for the `adiabatic expansion phase', where each SN can remain adiabatic for a timescale of $t_{\rm adi}$. This timescale is set to be $10^6$ yr. We adopt a fixed canonical stellar initial mass function (IMF) proposed by \citet{kroupa2001}, which in turn determines the chemical evolution, SN feedback, and dust formation and evolution.

Chemical enrichment through star formation and metal ejection from SNIa, SNII, and AGB stars is self-consistently included in the chemodynamical code. The code explicitly evolves 11 chemical elements (H, He, C, N, O, Fe, Mg, Ca, Si, S, and Ba) in order to predict both chemical abundances and dust properties. There is a time delay between the epoch of star formation
and those  of supernova explosions and the commencement of AGB phases (i.e., non-instantaneous recycling of chemical elements). We adopt the nucleosynthesis yields of SNe II and Ia from 
\citet{tsujimoto95} and AGB stars from \citet{vdh97} in order to estimate chemical yields.  Dust can grow through accretion of existing metals onto dust grains with a timescale of $\tau_{g}$. Dust grains can be destroyed though supernova blast waves in the ISM of galaxies and the destruction process is parameterised by the destruction time scale ($\tau_{\rm d}$). As discussed in B13, we consider models with $\tau_{\rm g}=0.25$ Gyr and $\tau_{\rm d}=0.5$ Gyr.  Despite the fact that the adopted code contains the above features related to chemical evolution and dust evolution, we do not investigate these properties in the present study.

The temperature ($T_{\rm g}$), hydrogen density ($\rho_{\rm H}$),  dust-to-gas ratio ($D$) of a gas particle, and the strength of the FUV radiation field ($\chi$) around the gas particle are calculated at each time step so that the fraction of molecular hydrogen ($f_{\rm H_2}$) for the gas particle can be derived based on the ${\rm H_2}$ formation/destruction equilibrium conditions. The SEDs of stellar particles around each $i$-th gas particle  (thus ISRF) are first estimated from ages and metallicities of the stars by using stellar population synthesis codes for a given IMF (e.g., \citealt{bruzual2003}). Then the strength of the FUV-part of the ISRF is estimated from the SEDs so that $\chi_i$ can be derived for the $i$-th gas particle. Based on $\chi_i$, $D_i$, and  $\rho_{\rm H, \it i}$ of the gas particle, we can derive $f_{\rm H_2, \it i}$ (see Figure 1 in B13). Thus each gas particle has $f_{\rm H_2, \it i}$, metallicity ([Fe/H]), and gas density.  The total dust, metal, and ${\rm H_2}$ masses are estimated from these properties.

\section{High resolution images of representative S0 models} \label{sec:hires}

As explained in Section \ref{sec:resref}, we chose the resolution of our images and velocity maps to match the spatial resolutions of IFS surveys. Here we show three representative S0 models at high resolution in order to reveal any structure which is present in the simulations but is concealed in the low resolution images that we supply to the CNNs (such as those of Figures \ref{fig:den} and \ref{fig:vel}).  S0 simulations for the isolated, tidal, and merger pathways are given in Figures \ref{fig:appendix_iso}, \ref{fig:appendix_tidal}, and \ref{fig:appendix_merger}, respectively. Each of these S0s originated from a Spiral galaxy of Model C (see Table \ref{tab:ic}).  A number of distinguishing features are apparent in the surface density and velocity maps, such as the magnitude of out-of-plane motions, the presence of a bar, and the amount of randomness in particle location and velocity as discussed in Section \ref{sec:random}.

\begin{figure}
   \begin{center}
   \includegraphics[width=0.49\textwidth]{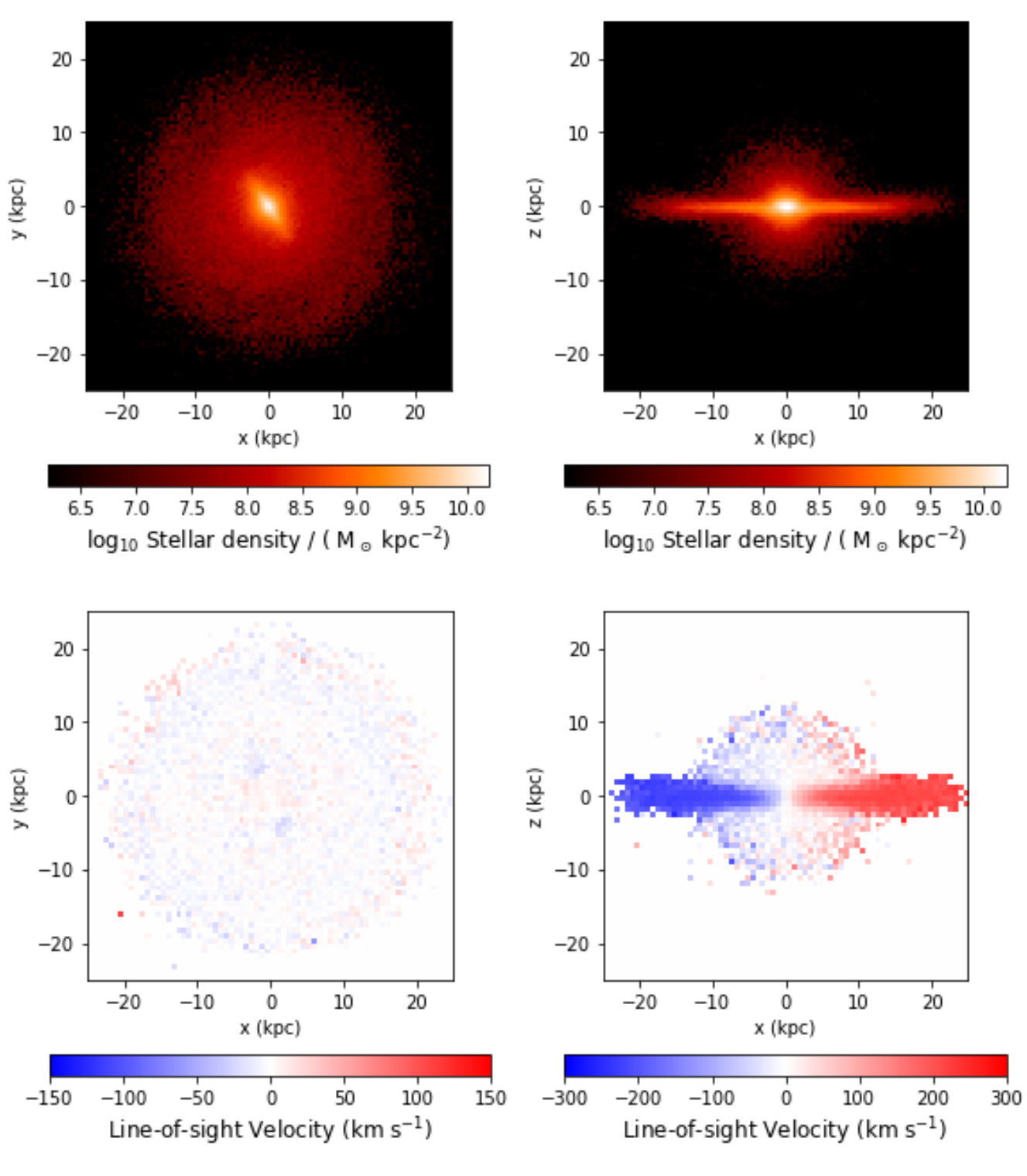}
   \caption{Representative S0 model of the isolated formation pathway, showing the stellar density in the face-on and edge-on orientations (top left and top right, respectively), as well as the two-dimensional velocity map shown face-on and edge-on (bottom left and bottom right, respectively).  This model originated from a Spiral galaxy of Model C (Table \ref{tab:ic}) with the Toomre $Q$ parameter set to 0 as explained in Section \ref{sec:ic}.}
   \label{fig:appendix_iso}
   \end{center}
\end{figure}

\begin{figure}
   \begin{center}
   \includegraphics[width=0.49\textwidth]{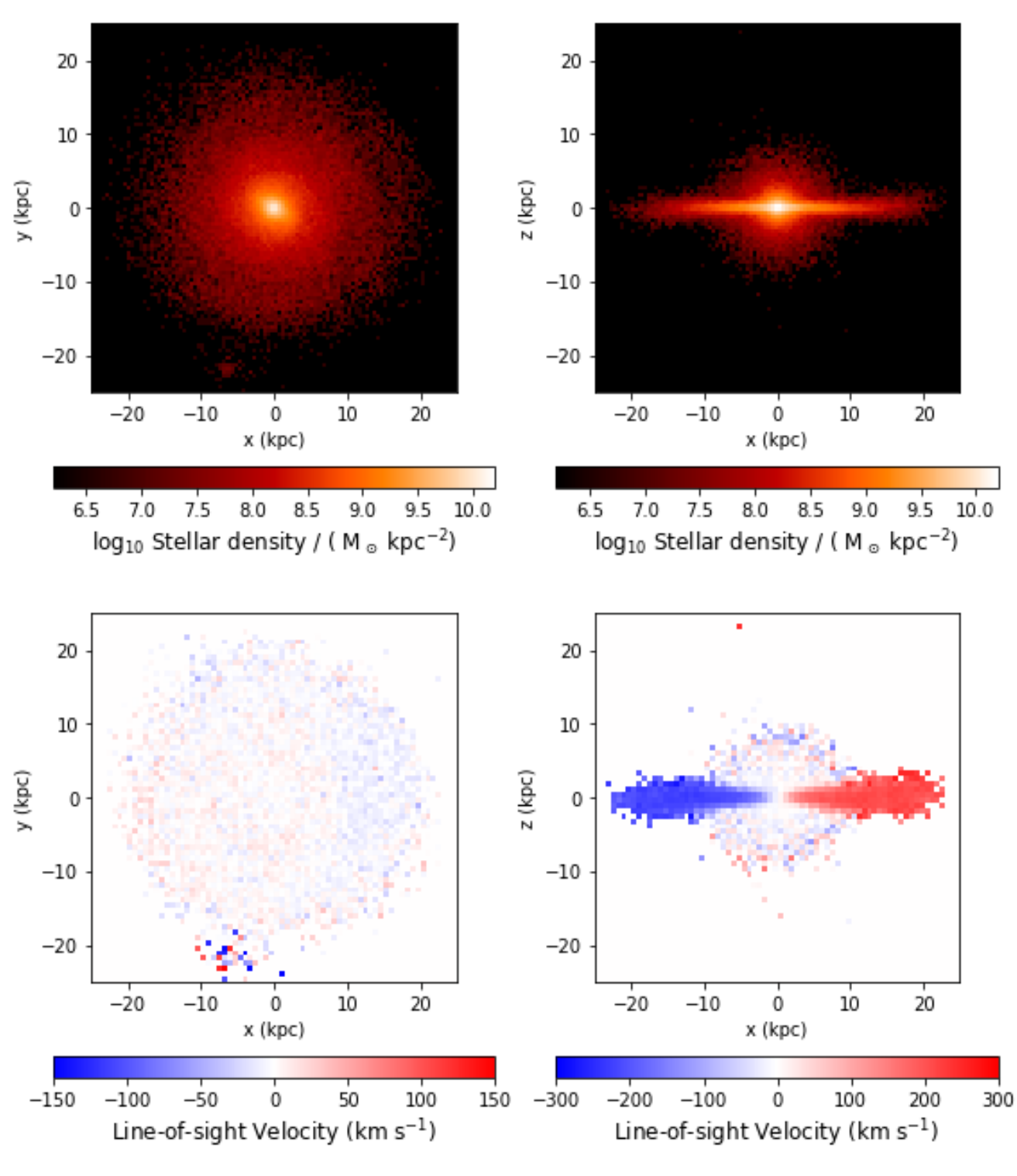}
   \caption{Representative S0 model of the tidal formation pathway, showing the stellar density in the face-on and edge-on orientations (top left and top right, respectively), as well as the two-dimensional velocity map shown face-on and edge-on (bottom left and bottom right, respectively).  This model originated from a Spiral galaxy of Model C (Table \ref{tab:ic}) placed in orbit around a group halo of total mass $2.5 \times 10^{13} ~ \Msun$. }
   \label{fig:appendix_tidal}
   \end{center}
\end{figure}

\begin{figure}
   \begin{center}
   \includegraphics[width=0.49\textwidth]{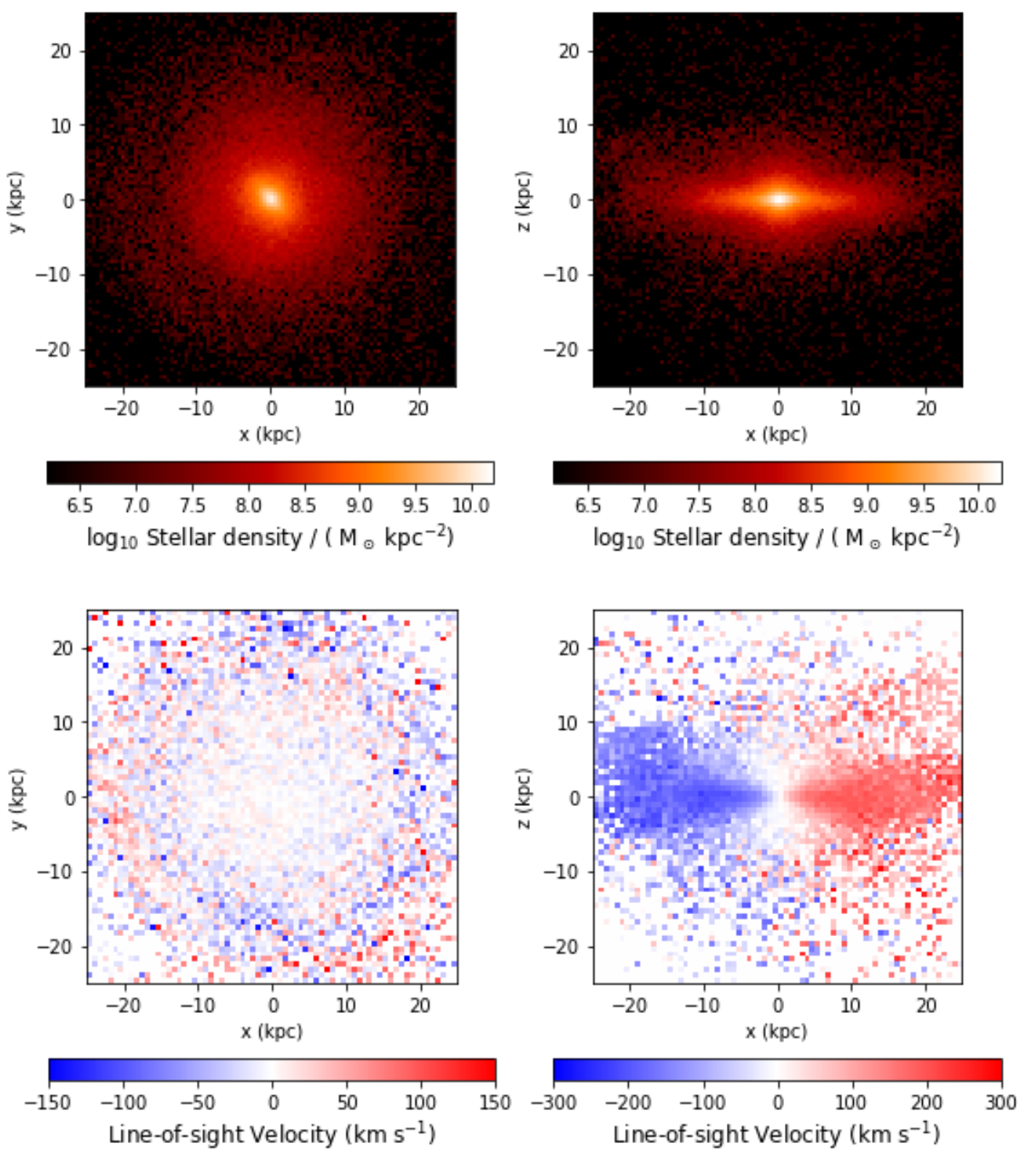}
   \caption{Representative S0 model formed by a merger, showing the stellar density in the face-on and edge-on orientations (top left and top right, respectively), as well as the two-dimensional velocity map shown face-on and edge-on (bottom left and bottom right, respectively).  This model originated from a Spiral galaxy of Model C (Table \ref{tab:ic}) and a merging satellite having a mass ratio of 0.11.}
   \label{fig:appendix_merger}
   \end{center}
\end{figure}

}

\end{document}